\documentclass[iop]{emulateapj}

\usepackage{rotating}
\usepackage{amsmath}
\usepackage{hyperref}
\usepackage{threeparttable}
\usepackage{booktabs, dcolumn}
\usepackage{amssymb}
\usepackage{pifont}

\newcommand{\beq}{\begin{equation}}
\newcommand{\eeq}{\end{equation}}
\newcommand{\xmark}{\ding{55}}

\def\earth{\hbox{$\oplus$}}
\def\arcmin{\hbox{$^\prime$}}
\def\arcsec{\hbox{$^{\prime\prime}$}}

\def\deg{\hbox{$^\circ$}}
\newcommand{\lsim}{\ \raise
-2.truept\hbox{\rlap{\hbox{$\sim$}}\raise5.truept\hbox{$<$}\ }}
\newcommand{\gsim}{\ \raise
-2.truept\hbox{\rlap{\hbox{$\sim$}}\raise5.truept\hbox{$>$}\ }}
\newcommand{\simsim}{\ \raise
-2.truept\hbox{\rlap{\hbox{$\sim$}}\raise5.truept\hbox{$\sim$}\ }}

\renewcommand{\arraystretch}{1.2}

\slugcomment{To appear in ApJ}
\shorttitle{IN-SYNC Orion}
\shortauthors{Da Rio et al. 2014}

\begin{document}

\title{IN-SYNC IV - The Young Stellar Population in the Orion A Molecular Cloud}

\author{
Nicola~Da Rio$^1$,
Jonathan~C.~Tan$^{1,2}$,
Kevin~R.~Covey$^3$,
Michiel~Cottaar$^4$,
Jonathan~B.~Foster$^5$,
Nicholas~C.~Cullen$^1$,
John~J.~Tobin$^6$,
Jinyoung~S.~Kim$^7$
Michael~R.~Meyer$^8$,
David~L.~Nidever$^9$,
Keivan~G.~Stassun$^{10}$,
S.~Drew~Chojnowski$^{11}$,
Kevin~M.~Flaherty$^{12}$,
Steve~Majewski$^{11}$,
Michael~F.~Skrutskie$^{11}$,
Gail~Zasowski$^{11,13,14}$,
\and Kaike~Pan$^{15}$}

\affil{ \\
$^1$Department of Astronomy, University of Florida, Gainesville, FL 32611, USA. \\
$^2$Department of Physics, University of Florida, Gainesville, FL 32611, USA. \\
$^3$Department of Physics \& Astronomy, Western Washington University, Bellingham, WA 98225, USA. \\
$^4$Department of Clinical Neurosciences, University of Oxford, Oxford, United Kingdom. \\
$^5$Yale Center for Astronomy and Astrophysics, Yale University New Haven, CT 06520, USA. \\
$^6$Leiden Observatory, NL-2333CA Leiden, The Netherlands
$^7$Steward Observatory, University of Arizona, Tucson, AZ 85721, USA. \\
$^8$Institute for Astronomy, ETH Zurich, Zurich, Switzerland. \\
$^9$Department of Astronomy, University of Michigan, Ann Arbor, MI 48109, USA.\\
$^{10}$Department of Physics \& Astronomy, Vanderbilt University, Nashville, TN 37235, USA.\\
$^{11}$Department of Astronomy, University of Virginia, Charlottesville, VA 22904, USA. \\
$^{12}$Astronomy Department, Wesleyan University, Middletown, CT 06459, USA.\\
$^{13}$Department of Astronomy, The Ohio State University, Columbus, OH 43210, USA.\\
$^{14}$Center for Cosmology and Astro-Particle Physics, The Ohio State University, Columbus, OH 43210, USA. \\
$^{15}$Apache Point Observatory and New Mexico State University, P.O. Box 59, Sunspot, NM, 88349-0059, USA.}

\email{ndario@ufl.edu}


\begin{abstract}
We present the results of the SDSS APOGEE \emph{INfrared Spectroscopy of Young Nebulous Clusters program} (IN-SYNC) survey of the Orion A molecular cloud. This survey obtained high resolution near infrared (NIR) spectroscopy of about 2700 young pre-main sequence stars throughout the region, acquired across five distinct fields spanning 6\deg\ field of view (FOV). With these spectra, we have measured accurate stellar parameters ($T_{\rm eff}$, $\log g$, $v\sin i$) and extinctions, and placed the sources in the Hertzsprung-Russel Diagram (HRD). We have also extracted radial velocities for the kinematic characterization of the population. We compare our measurements with literature results for a sub-sample of targets in order to assess the performances and accuracy of the survey. Source extinction shows
evidence for dust grains that are larger than those in the diffuse interstellar medium (ISM): we estimate an average $R_V=5.5$ in the region.
{ Importantly, we find a clear correlation between HRD inferred ages and spectroscopic surface-gravity inferred ages. This clearly indicates a real spread of stellar radii at fixed temperature, and together with additional correlations with extinction and with disk presence, strongly suggests a real spread of ages large than a few Myr.}
Focussing on the young population around NGC1980 / $\iota$ Ori, which has previously been suggested to be a separate, foreground, older cluster, we confirm its older ($\sim5$~Myr) age and low $A_V$, but considering that its radial velocity distribution is indistinguishable from the Orion A's population, we suggest that NGC1980 is part of Orion A’s star formation activity. Based on their stellar parameters and kinematic properties, we identify 383 new candidate members of Orion A, most of which are diskless sources in areas of the region poorly studied by previous works. 
\end{abstract}



\section{Introduction}
\label{section:introduction}
Young stellar populations are the immediate product of converting gas in molecular clouds into stars, and therefore provide critical information on how star formation progresses over time and space. Despite recent advances that have been made in this field, including observational and theoretical work focussing on the different stages of star formation, a number of mechanisms remain unclear. It is still debated if star formation is a fast, dynamic process \citep{elmegreen2000,elmegreen2007,hartmann2007} or slow
\citep{tan2006,dario2014a}, proceeding for at least several dynamical timescales. In this context, the systematic, unbiased assessment of stellar ages and age spreads in young pre-main sequence (PMS) populations is fundamental \citep{dario2010b,dario2014a}, though not easy given our poor understanding of PMS evolution and its dependence of the earlier protostellar properties and accretion histories \citep{baraffe2009,baraffe2012}.
The star formation efficiency, both per free-fall time $\epsilon_{\rm ff}$ \citep{tan2006,dario2014b}, or on an absolute scale, as the fraction of parental gas converted into stars, and especially their environmental dependence are not fully understood. { These fundamental properties are also important factors in the survivability of a young population as a bound cluster \citep{ladalada2003,baumgardt2007,kruijssen2012}}. This is further complicated considering that observational evidence has shown that star formation typically occurs in filamentary structures which can show significant substructure at early phases \citep[e.g.,][]{andre2013,hacar2013,henshaw2014}. Such structures may then also merge in relatively short timescales, erasing the information on their initial morphology \citep{parker2012,parker2014b,dario2014b,jaehnig2015}.

The upcoming GAIA mission will help constrain the kinematic properties of nearby young populations through accurate proper motions and parallactic distances, however it will not perform well for young clusters still embedded in their parental material. Similarly, most ground-based radial velocity surveys have adopted optical spectroscopy \citep{tobin2009,gilmore2012,jeffries2014}. Recently, the IN-SYNC survey used the capabilities of the SDSS-III  Apache Point Observatory Galactic Evolution Experiment (APOGEE) to obtain multi-object near infrared (NIR) high-resolution spectra to obtain stellar parameters and radial velocities in young clusters. The first part of this survey covered the Perseus cloud, through their young clusters IC348 \citep[][, hereafter Paper~I and II]{cottaar2014,cottaar2015} and NGC1333 \citep[][, hereafter Paper~III]{foster2015}. In these studies we have demonstrated the abilities of APOGEE to study the stellar properties in young clusters, assess their age spread, and compare the kinematic properties of young stars with those of the remaining gaseous material. We have found evidence for a supervirial stellar population in IC348, whereas in NGC1333 the stars are in agreement with virial velocities, but the diffuse and dense gas have significantly different velocity dispersion. { As for the comparison between stellar and gas kinematics, Paper~II found that stars in NGC1333 show a similar velocity dispersion than the diffuse gas, whereas the young dense cores in the region have much slower, subvirial, motions.  In a second phase of the IN-SYNC collaboration we have extended our survey to the Orion A molecular cloud.}

The Orion molecular complex is a large structure, at a distance of $\sim 400$~pc \citep{menten2007,schlafly2014}, which has sustained star formation in different locations for over 15~Myr \citep{blaauw1964,blaauw1991,muench2008}. The oldest young populations, such as $\lambda\  Ori$, $\sigma\  Ori$ and $25\ Ori$ are mostly disassociated from the molecular material. Two main regions of dense molecular gas, referred to as Orion A and Orion B, host young (few Myr) stellar populations with ongoing star formation \citep{megeath2012,lombardi2011,lombardi2014,stutz2013,stutz2015}. Of these two, the 40~pc long Orion A filament has the highest star formation rate. The densest region in Orion A, the Orion Nebula Cluster (ONC), is the closest site of active massive star formation, where young  stars span the  mass spectrum up to $\sim 40~M_{\odot}$ \citep{grellmann2013}. Sparser populations are present to the north, in the upper sword and NGC1977, and to the south, with NGC1980, and L1641. All these populations have been long studied through photometric and spectroscopic studies \citep{hillenbrand1997,hillenbrand-hartmann1998,dario2010a,dario2012,robberto2010,robberto2013,hsu2012,hsu2013,fang2009,fang2013}, at longer wavelengths \citep{megeath2012,lombardi2014} and in the X-rays \citep{getman2005,pillitteri2013}. The PMS population throughout the region shares approximately the same age of a few Myr \citep{dario2010b,fang2013}, with the exception of the region around $\iota$ Ori, to the immediate south of the ONC, where \citet{alves2012} and \citet{bouy2014} claim that the population is slightly older and possibly in foreground, due to the low average $A_V$ of the members. The initial mass function (IMF) is found to be compatible with the \citet{kroupa2001} in the ONC, although is found to be deficient in the substellar regime \citep{dario2012}. In L1641, however, \citet{hsu2013} note that despite the large number of PMS members in this region (exceeding $10^3$ stars), no massive stars are present. This would suggest an upper truncation of the IMF, possibly due to lower density conditions.

Kinematic studies based on radial velocities have been so far limited to the north part of the cloud, with a field of view of $2\deg$ in declination. \citet{furesz2008} and \citet{tobin2009} collected velocities through optical high resolution spectroscopy for over 1600 sources. They measured an average velocity dispersion $\sigma_r\sim2.3$~km s$^{-1}$, consistent with proper motion measurements in the ONC \citep{jones-walker1988}. They also noted that stars in this region are systematically blueshifted by $\sim 1$ km s$^{-1}$ relative to the local gas.

In this work we extend over previous studies and present the results from the IN-SYNC Orion survey, including the study of the stellar population, extinction law, ages and age spreads and new memberships from a combination of indicators. In a subsequent paper we will focus on the APOGEE radial velocities to characterize the kinematic status of the entire region, in comparison with the molecular gas.

In Section \ref{section:data} we present the data and extraction of stellar parameters. In Section \ref{section:stellar_parameters} we compare the stellar parameters we derived with literature results for a subset of targets, in order to asses the accuracy of our measurements, and estimate the incidence of contaminants. In Section \ref{section:ages} we explore the age and age spread properties of the entire population. Last in Section \ref{section:membership} we utilize our measurements to recover new candidate members from sources previously unclassified.

\section{The Data}
\label{section:data}
Observations were carried out in December 2013 and January 2014, with the APOGEE spectrograph \citep{wilson2010,wilson2012} on the Sloan 2.5 m telescope \citep{gunn2006}. APOGEE is a fiber-fed multi object infrared spectrograph, operating in $H$-band in the range $1.5\mu m \lesssim \lambda \lesssim 1.6\mu m$, capable of obtaining spectra of up to 320 sources simultaneously on a corrected field of view of $\sim 7$ square degrees, and with a resolution $\lambda/\Delta \lambda\simeq22,500$. More information on this instrument, the standard observing procedures, the data reduction pipeline and the resulting spectra and catalogs can be found on the SDSS-III Web site (\url{http://www.sdss3.org}) and in the APOGEE technical papers \citep{zasowski2013,nidever2015,holtzman2015}.

\begin{figure*}
\epsscale{1.1}
\plotone{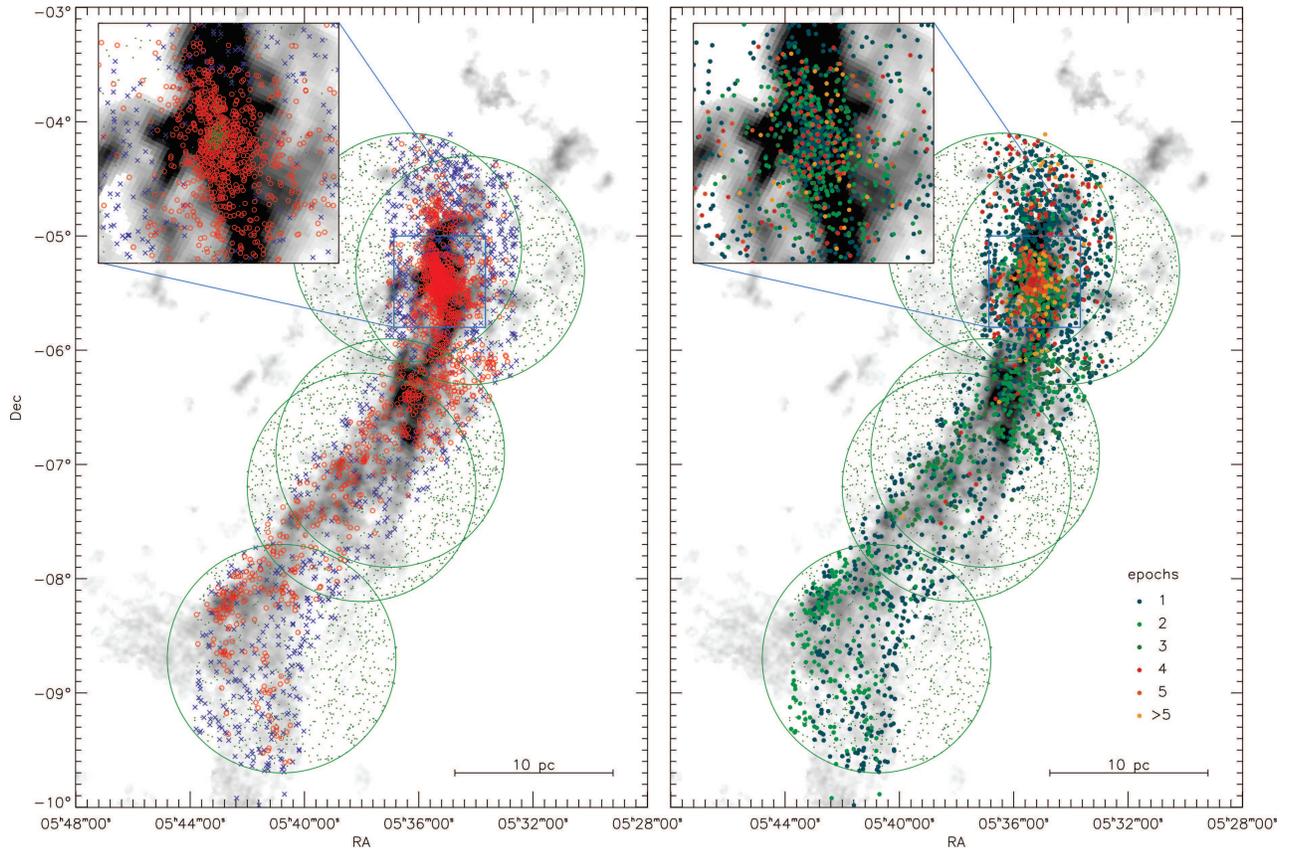}
\caption{The IN-SYNC Orion targets overplotted on the $^{13}$CO map from \citet{nishimura2015}. The left panel highlights the previously known young members from the literature (red circles) as opposed to targets of unknown memberships (blue). The right panel shows the targets color coded according to the number of epochs in which each star has been observed. Dots indicate sources with $H<12.5$ not targeted by our program. The green circles indicate the 2$\deg$ FOVs, smaller than the full APOGEE FOV, to which we restricted our 5 pointing positions. \label{figure:radec_over_13co}}
\end{figure*}

\begin{figure}
\epsscale{1.1}
\plotone{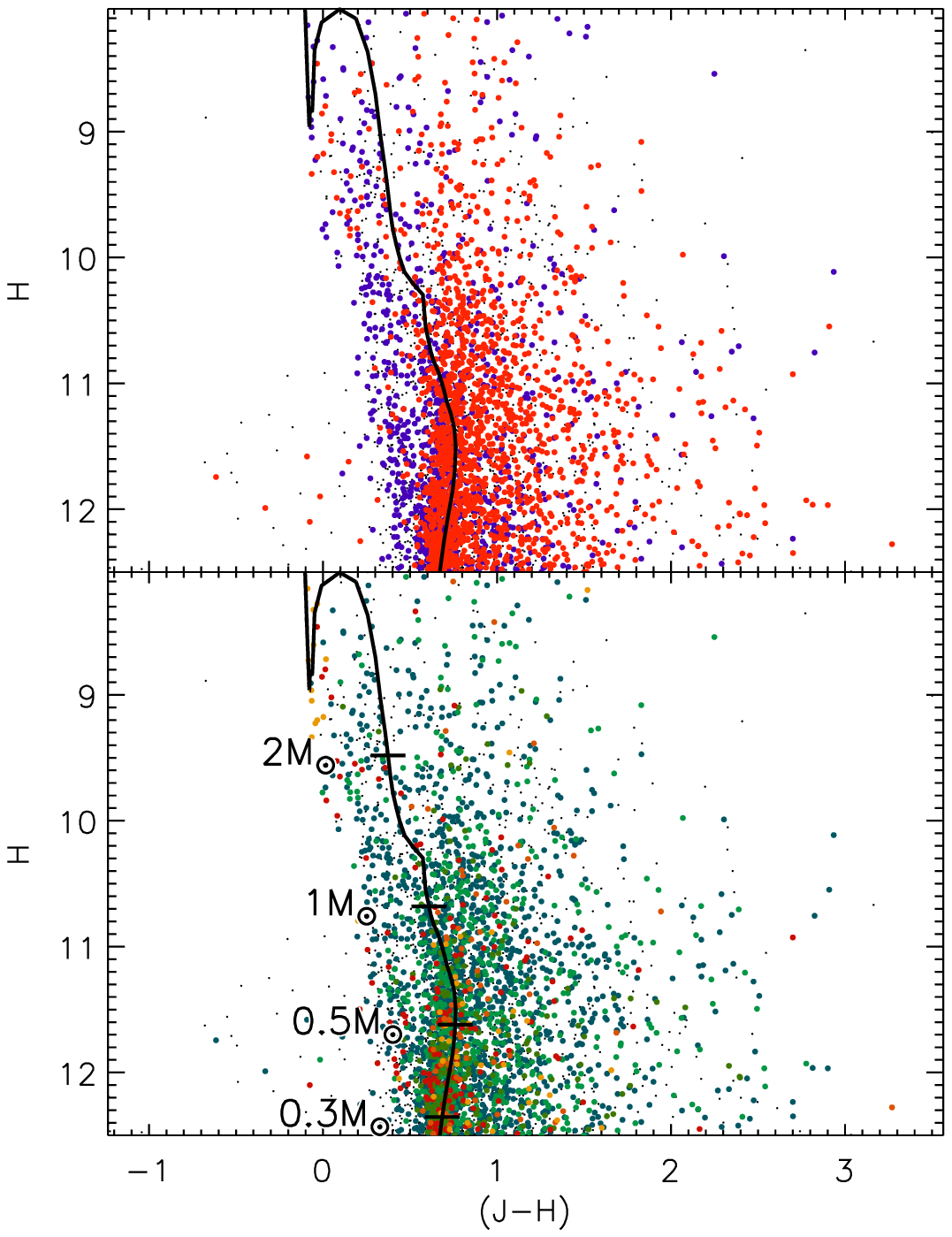}
\caption{2MASS CMDs of the selected targets. Top panel shows known members (red) and sources with unknown membership (blue), while the bottom panel indicates the number of epochs as in Figure \ref{figure:radec_over_13co}. The solid line is a 2.5~Myr isochrone from \citet{siess2000}.  \label{figure:cmds}}
\end{figure}


15 APOGEE plates, on 5 positions in the sky (see Table \ref{table:plates}), have been designed to cover the Orion A region as shown in Figure \ref{figure:radec_over_13co}. The area covered has been isolated to match the Spitzer survey of young stellar objects described by \citet{megeath2012}, which in turn follows the distribution of the molecular gas as traced by $^{12}$CO \citep{bally1987}. Specifically, the more populated two northernmost positions have been covered with 5 plates each, and the rest of the filament with the remaining five. Targets have been selected among {\em Two Micron All-Sky Survey} (2MASS) sources with $H<12.5$, using an automatic procedure. This aimed at maximizing the number of allocated fibers per plate while also accounting for priority rankings among the individual sources. To this end, we developed an automatic stochastic fiber allocation method. For each plate, a random observable source is considered. Then, based on targeting probability defined for that star, it is randomly kept, allocating a fibers to it, or rejected. If rejected, the star will be still available for subsequent iteration of the method, which continues until until all available fibers have been allocated.
The targeting probabilities have been tuned after several attempts to achieve a sensible overall targeting.
Known members from optical studies, in the ONC \citep{dario2010a,dario2012} or in L1641 \citep{fang2009,fang2013,hsu2012,hsu2013}, with available stellar parameters were given the highest priority. Additional sources  with IR-excess \citep{megeath2012} or X-ray emission \citep{getman2005,pillitteri2013} were a lower probability. Stars in the crowded core of the ONC were also ranked higher, to compensate for the intrinsic difficulty in targeting them due to the large exclusion radius ($\sim 1.2\arcmin$) from fiber collision on the APOGEE plates. This involved multiplying the targeting population probabilities by a factor which continuously depends on the number of neighbors in the vicinity of each source.
Unlike the IN-SYNC Perseus survey, in Orion we prioritized the number of individual sources observed over the repetition of the same sources at different epochs. Thus, targeting probability of each star is reduced 50 times if the star was already targeted by a different plate, or only 10 times for sources at the faint end of the APOGEE luminosity range (11.5$<H<$12.5) to improve their signal to noise ratio (SNR).
All the plates were filled with 320 targets, plus additional sky and telluric fibers. Figure \ref{figure:radec_over_13co} shows the spatial distribution of the targeted sources, superimposed on a $^{13}$CO map from \citet{nishimura2015}, highlighting the previously known members, as well as the number of epochs observed for the same star.
Hereafter we will refer to sources as ``known members'' if they satisfy any of the following criteria:
\begin{itemize}
\item IR excess, from the 2MASS and Spitzer survey of \citep{megeath2012}.
\item X-ray sources matching optical or IR photometry, in the ONC from the COUP survey \citep{getman2005}, Chandra observations in the ONC flanking fields \citep{ramirez2004}, and XMM-Newton observations in L1641 \citep{pillitteri2013}.
\item Sources in L1641 with spectroscopic evidence of youth, either Lithium abundance or H$\alpha$ excess \citep{fang2009,fang2013}, as well as additional candidate members from \citet{hsu2012}
\item Additional ONC sources with HRD position consistent with the PMS sequence of the cluster \citet{dario2010a,dario2012}.
\end{itemize}

In Section \ref{section:membership} we will extend the sample of candidate members based on our data; however in all material preceeding Section \ref{section:membership} we will refer to these previously identified members as ``known members''.

Figure \ref{figure:cmds} shows our targets in a 2MASS $JH$ color-magnitude diagram (CMD). Including telluric fibers in our FOV, we collected 4828 spectra of 2691 individual sources, 1704 of which were previously identified members from the literature.
Thanks to our forced priority in targeting previously known members, were able to cover most of them (80--90\%) within the entire region down to our selection limit $H<12.5$, with the sole exception of the core of the ONC, where due to crowding the coverage drops to about a half.

\begin{deluxetable}{cccc}[b]
\tablecaption{APOGEE plate centers\label{table:plates}}
\tablehead{\colhead{center} & \colhead{$\alpha$} & \colhead{$\delta$} & \colhead{number } \\
\colhead{} & \colhead{(J2000)} & \colhead{(J2000)} & \colhead{of plates} }
\startdata
A & $05^h36^m24^s$ & $-05\deg06$\arcmin 00\arcsec & 5 \\
B & $05^h34^m12^s$ & $-05\deg18$\arcmin 00\arcsec & 5 \\
C & $05^h37^m00^s$ & $-06\deg54$\arcmin 00\arcsec & 2 \\
D & $05^h38^m00^s$ & $-07\deg12$\arcmin 00\arcsec & 1 \\
E & $05^h40^m48^s$ & $-08\deg42$\arcmin 00\arcsec & 2
\enddata
\end{deluxetable}

The raw data have been processed through the APOGEE pipeline described in \citet{nidever2015}, which corrects for instrumental effects and performs wavelength and flux calibration for each spectrum. Stellar parameters have been then derived by us by comparison of our observations with BT-Settl synthetic spectra \citep{allard2011}. We adopted a Markov Chain Monte Carlo (MCMC) algorithm to obtain the best fit, leaving $T_{\rm eff}$ and $\log g$ free and constraining [M/H]=0 as appropriate for the region \citep{dorazi2009}. Radial velocity $v_r$ and projected rotational velocity $v\sin i$ have also been left free; for the latter the models have been convolved with the rotational broadening profile from \citet{gray1992}. Last, a variable amount of flat flux excess has been added to the spectra and left as a free parameter, to emulate the amount of veiling -- non-photospheric flux excess originating from either circumstellar disk thermal emission, or from the accretion flow. Further details on the fitting procedure are presented in Paper~I. As highlighted in that work, analyzing the parameters derived in all the IN-SYNC clusters compared to values reported in the literature for the same sources, we found systematic departures in our $v_r$ and $T_{\rm eff}$ at the low-end of the $T_{\rm eff}$ scale, and attributed them to inaccuracies of the BT-Settl synthetic models. As a matter of fact, these models have been shown to deviate from empirical data in the range $3000K\lesssim T_{\rm eff} \lesssim 3500K$, both in the optical \citet{dario2010a} and in the NIR, specifically in the $H$-band \citep{scandariato2012,kopytova2013} where APOGEE operates.

Our MCMC algorithms naturally provides associated uncertainties for each fitted parameter. As described in Paper~I, by studying the scatter in the best fit parameters of the same star at different epochs, we found that this derived error represents an underestimate of the real error. A correction factor was applied to scale the MCMC error estimates to the empirical values empirically determined to depend on the $\chi^2$ value of each fit. Such discrepancy, which we corrected for in our Orion dataset as well, probably originates from the inaccuracies of the synthetic spectra to reproduce the real data.

\section{Stellar parameters}
\label{section:stellar_parameters}
We have compared our derived stellar parameters with estimates from the literature, where available for a subsample of sources. In the ONC, stellar censuses have been collected from optical spectroscopy \citep{hillenbrand1997,hillenbrand2013,dario2010a} and from narrow band photometry mapping temperature sensitive TiO bands in the M-type range \citep{dario2010a,dario2012}; radial velocities have been collected by \citet{sicilia-aguilar2005}, \citet{furesz2008} and \citet{tobin2009}. In L1641, optical spectroscopy was carried out by \citet{fang2009,fang2013} and \citet{hsu2012,hsu2013}, but no radial velocities have been collected for declinations $\delta<-6\deg$.

\subsection{Effective temperature}
\label{section:stellar_parameters-teff}
\begin{figure}
\plotone{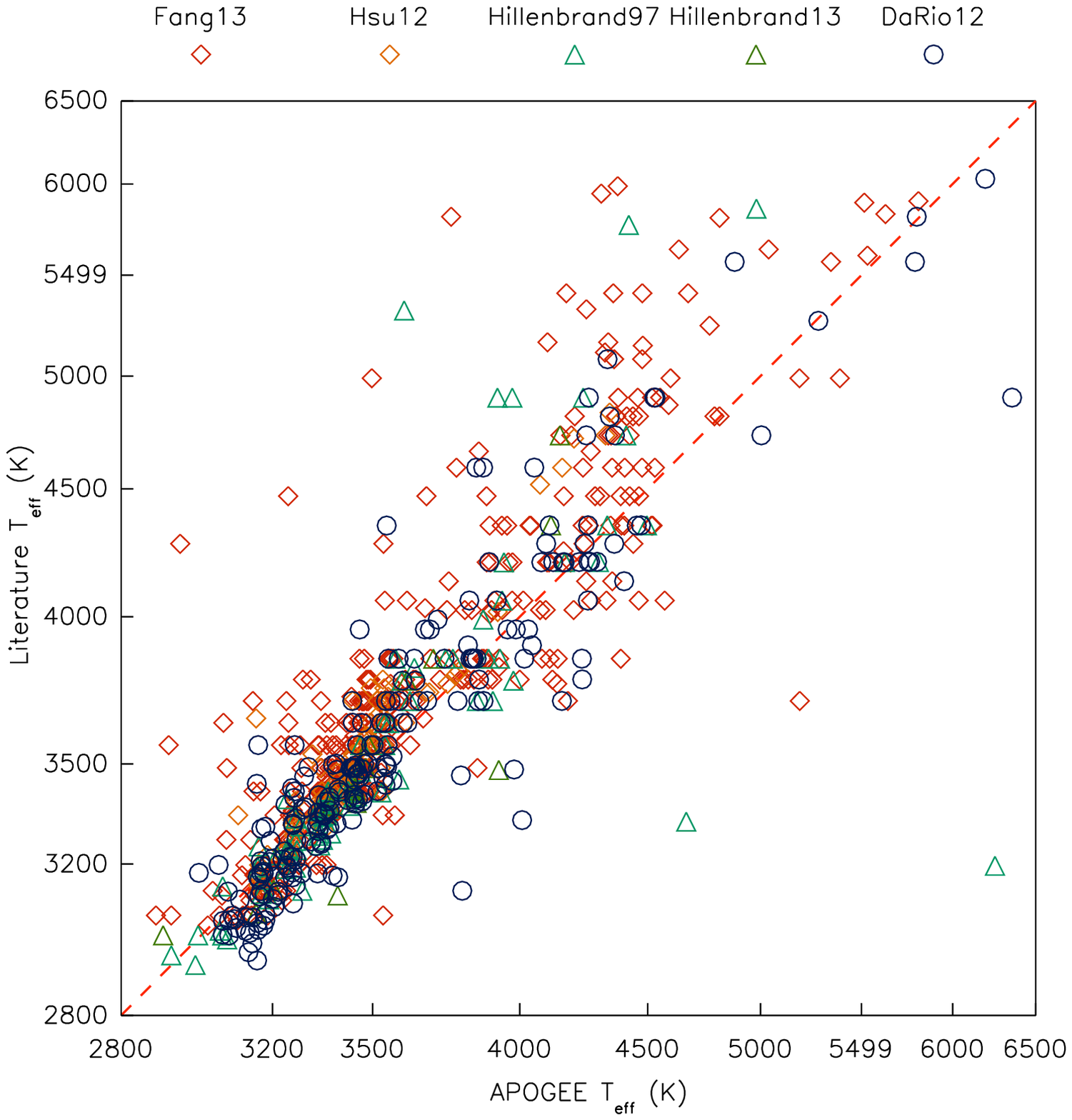}
\caption{Comparison between our APOGEE derived $T_{\rm eff}$ and values optical studies in the literature \citep{fang2013,hsu2012,hillenbrand1997,hillenbrand2013,dario2012}, respectively in L1641 (first two) and the ONC (remaining three). \label{figure:teff_comparison}}
\end{figure}

\begin{figure}
\plotone{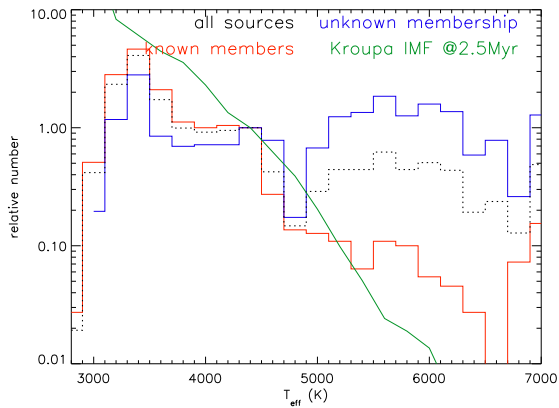}
\caption{Distribution of $T_{\rm eff}$ extracted from our APOGEE sample, as well as that limited to sources with known memberships or not. The green line is the predicted distribution for a (complete) Kroupa IMF assuming a \citet{siess2000} 2.5~Myr isochrone \label{figure:teff_histograms}}
\end{figure}

Figure \ref{figure:teff_comparison} shows the comparison of our derived effective temperatures ($T_{\rm eff}$) and previous estimates. As discussed in Paper~I for the Perseus survey, our APOGEE spectra are less accurate in deriving $T_{\rm eff}$ for intermediate temperatures ($T_{\rm eff}\gtrsim4000$K, due to the smaller amount of features in the $H$-band spectra compared to colder sources; this is evident from Figure \ref{figure:teff_comparison} as a large scatter in this $T_{\rm eff}$ range. However, at increasing $T_{\rm eff}$ the fraction of field contaminants increases (see below), such that most of the young Orion members are in the M-type range. Below $4000$K the agreement is much tighter, especially for ONC stars \citep{hillenbrand1997,hillenbrand2013,dario2012}, whereas the scatter increases for L1641 members. This indicates that the literature parameters in L1641 were less reliable than those for the ONC. The correlation is particularly tight when limiting to sources from \citet{dario2012}, with $T_{\rm eff}$ derived from narrow band photometry, with a RMS between new and old results of $\sim100$K. The RMS increases to $\sim140$K when comparing with optical spectroscopy from \citet{hillenbrand1997} and
\citet{hillenbrand2013}. In turn, \citet{hillenbrand2013} that when comparing independent $T_{\rm eff}$ values, the scatter was larger between spectroscopically derived parameters compared to narrow band values. Our results here therefore suggests that also our APOGEE $T_{\rm eff}$ are more accurate than moderate resolution optical spectra in the M-type mass range.

Last, Figure \ref{figure:teff_histograms} compares the $T_{\rm eff}$ distributions of previously known members with that of unknown membership sources, as well as the total number of APOGEE targets; the distributions have been normalized at 4400K. We also report the predicted distribution from a \citet{kroupa2001} initial mass function (IMF), assuming a 2.5~Myr isochrone from \citet{siess2000}. It is evident that at intermediate temperatures ($T_{\rm eff}\gtrsim5000$) the sample of sources with unknown memberships is overpopulated relative to the Galactic IMF. Such enhancement of intermediate-mass star has never been detected, for members, in the the region \citep{dario2010a,hsu2013}. We thus conclude that in this $T_{\rm eff}$ range, sources without confirmed memberships are mostly field contaminants, likely low- and intermediate-mass dwarfs. At lower $T_{\rm eff}$, the very low mass field dwarfs will be likely too faint to contribute a significant contamination to our sample. It is also noteworthy that even the sub sample of known members shows a somewhat excessive number of sources at warmer temperatures compared to what is expected from a standard IMF. This is due to an incompleteness bias, in that members with lower $T_{\rm eff}$ and higher $A_V$ will be more likely excluded from our sample because of our targeting luminosity threshold ($H<12.5$).

\subsection{Extinction}
\label{section:stellar_parameters-av}
\begin{figure*}
\epsscale{0.8}
\plotone{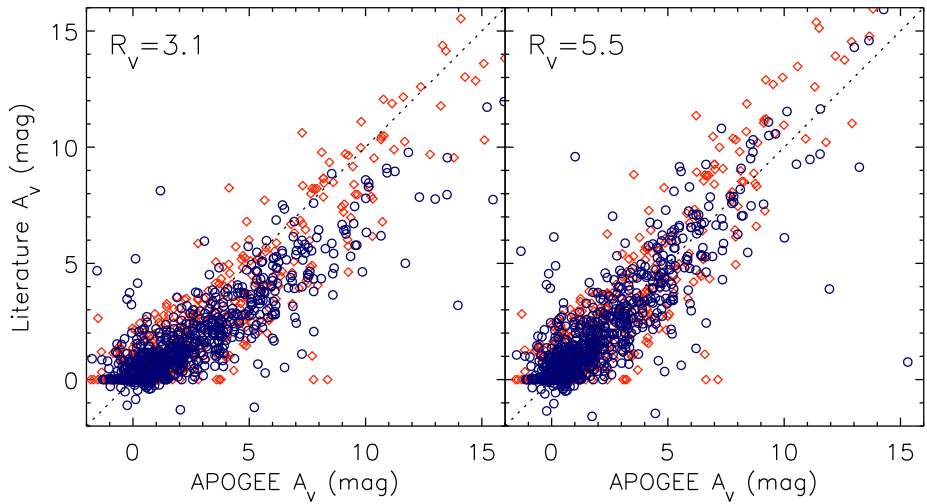}
\caption{Comparison between our APOGEE derived extinctions, and values from optical data in the literature, assuming $R_V=3.1$ (left panel) or $R_V=5.5$ (right panel) respectively, on both axes. Blue circles are for ONC sources, red diamonds for L1641 stars. \label{figure:av_comparison}}
\end{figure*}

We derived the extinction of our IN-SYNC targets from the knowledge of $T_{\rm eff}$ and the observed colors. Specifically, we compared the sources' observed $(J-H)$ colors with the intrinsic ones from the 2.5~Myr old semi-empirical isochrone from \citet{bell2014}, obtained applying an empirical correction to the PISA models \citep{tognelli2012} as converted into magnitudes assuming BT-Settl \citep{allard2011} synthetic spectra. To convert the reddening excess $E(J-H)$ to $A_V$, we assumed the reddening curve from \citet{cardelli1989} for a reddening parameter $R_V=3.1$ or $5.5$. We compared our derived $A_V$ with values from the literature \citep{dario2012,hillenbrand2013,fang2013}, which were all derived assuming the Galactic $R_V=3.1$. This comparison is shown in Figure \ref{figure:av_comparison}, where the left panel shows that our APOGEE $A_V$ values deviate systematically from the optically derived ones. Given the different wavelength ranges adopted to measure the reddening, this could indicate that a different reddening law is required. We thus corrected the literature extinctions as if they had been computed assuming $R_V=5.5$. This was accomplished considering the different color excesses used in each of those works to obtain $A_V$: $E(V-I)$ for \citet{hillenbrand1997,hillenbrand2013} and \citet{dario2010a}, $E(7500\AA-I)$ for the sources with medium band photometry from \citet{dario2012}, and a combination of SDSS $griz$ and 2MASS $IJ$ bands for L1641 sources in \citet{fang2013}, for which we adopted a wavelength range represented by the color term ($i-J$). Figure \ref{figure:av_comparison}, right panel, confirms that assuming $R_V=5.5$ allows the different $A_V$ estimates to be in better agreement.
From the left panel of Figure \ref{figure:av_comparison} it appears that, for the sources in L1641 the discrepancy between APOGEE and literature values of $A_V$ is more moderate compared to that of the ONC sources. This is because $A_V$ was estimated from the SED at longer wavelengths, thus less sensitive to changes in the reddening law than in the optical. Because of this, their correction needed in the assumption of $R_V=5.5$ (Figure \ref{figure:av_comparison}, right panel) is more modest.

Another possibility for the discrepancy in $A_V$ values assuming $R_V=3.1$ may be the contribution to the $JH$ photometry from disk emission. This emission would increase the $(J-H)$ color term leading us to overestimate the stellar reddening. We thus compared the distribution of sources in the diagram of Figure \ref{figure:av_comparison} for different ranges of $(K-8\mu m)$ infrared slopes, but found no dependence at all on the infrared slope. Thus we concluded that NIR disk excess emission does not bias our $A_V$ estimates.

An increase of $R_V$ compared to the average Galactic value in star-forming regions at increasing densities, indicating grain growth, is not surprising and has been reported before \citep{johnson1968,costero1970,mccall2004,keto2008,foster2013,demarchi2014}. In the ONC, however, \citet{dario2010a} found $R_V=3.1$ more suitable to reproduce the distribution of sources in the $BVI$ two-color diagram for the lightly embedded sources surveyed in that study ($A_V< 5$). It is thus likely that the reddening parameter increases along the line of sight, from near Galactic values at the edge of the system to larger grains at higher optical depth. This is consistent with the fact that the volume density of molecular material, at least in the ONC \citep{dario2014b} increases along the line of sight, with most of the mass and density concentrated at the far end of the stellar cluster. Thus the more heavily embedded members, which are affected by a flatter reddening law, are engulfed in denser molecular material than the lightly embedded population.

\subsection{Surface gravity}
\label{section:stellar_parameters-logg}

\begin{figure*}
\epsscale{0.8}
\plotone{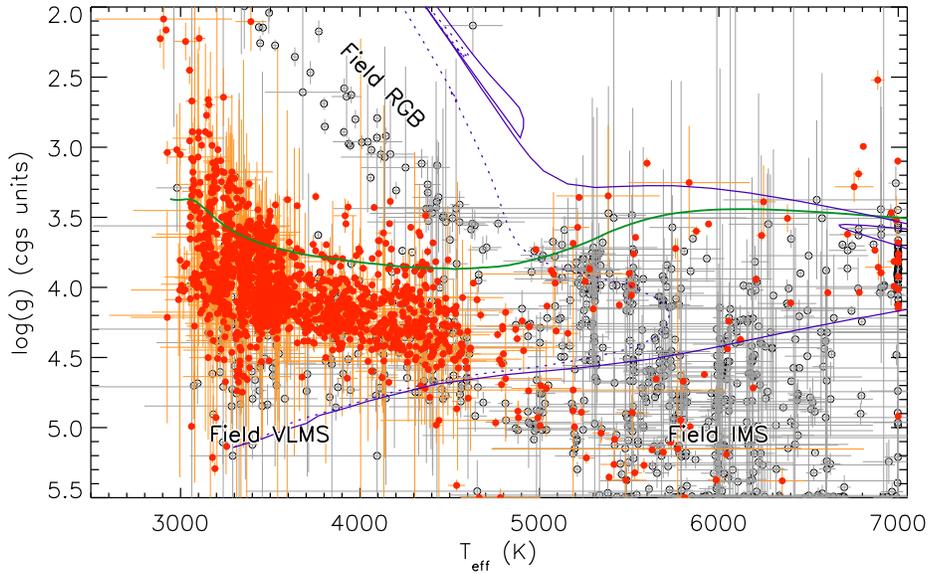}
\caption{Surface gravity versus $T_{\rm eff}$ plot. The red symbols indicate known members, the open circles sources of unknown membership and non-members. The solid green line is a 2.5~Myr isochrone from \citet{siess2000}; the blue solid and dotted lines are evolved isochrones of 10 and 1 Gyr of age from Padova evolutionary models \citep{marigo2008}  \label{figure:teff_logg}}
\end{figure*}

Figure \ref{figure:teff_logg} shows the surface gravity $\log g$ vs $T_{\rm eff}$ plot for all our targets, highlighting previously known members from different indicators in the literature. These include HRD position membership from \citet{dario2012}, X-ray sources \citet{getman2005}, IR-excess sources \citet{megeath2012}, and sources showing either undepleted lithium or H$\alpha$ excess \citep{fang2009,fang2013,hsu2012}. Most of the known members have cool $T_{\rm eff}$, as expected from the IMF peaking in the very low mass star (VLMS) range. The $\log g$ values are lower than those predicted from dwarfs, compatible with a PMS status of the sources. Although the diagram shows systematic departures from the expected locus, indicated for example by the \citet{siess2000} 2.5~Myr isochrone that best fits the ONC in the HRD, our Paper~I has already shown that relative cluster ages are well detected from APOGEE derived $\log g$ as an increasing sequence from NGC~1333, to IC~348, to the Pleiades.

Overall, the $\log g$--$T_{\rm eff}$ sequence for known Orion members remains relatively tight for below 4500~K, with a standard deviation of the order of 0.5~dex. The scatter -- both for candidate members and unknown sources increases at hotter temperature, consistent with the overall poorer performances of APOGEE in this regime (see also Section \ref{section:stellar_parameters-teff}).

A clear diagonal feature, { composed by sources without confirmed membership} is evident in Figure \ref{figure:teff_logg} extending from the upper left ($T_{\rm eff}\sim2500\ {\rm K}$ and $\log g\sim2$) to $T_{\rm eff}\sim5000\ {\rm K}$ and $\log g\sim4$ where it begins to blend with the field population. Although poorly matched by evolutionary model prediction, we safely identify it as a red giant branch of contaminating field sources. Additional field sources may be located at low $T_{\rm }$ at high surface gravities typical of main sequence dwarfs. Last, at intermediate $T_{\rm eff}$ the large majority of sources with no membership estimates show high $\log g$ values, consistent with being mostly field dwarfs, as already discussed in Section \ref{section:stellar_parameters-teff}.

\subsection{Radial velocity}
\label{section:stellar_parameters_rv}

\begin{figure}
\epsscale{1.2}
\plotone{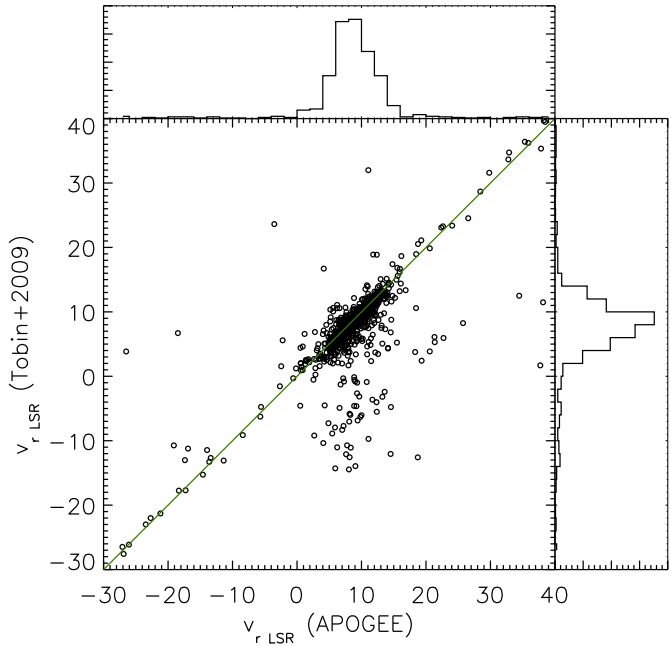}
\caption{Comparison with our derived radial velocity $v_r$ and literature estimates from optical spectroscopy \citep{tobin2009}.   \label{figure:rv_comparison}}
\end{figure}

Figure \ref{figure:rv_comparison} shows a comparison of our APOGEE-derived radial velocities $v_r$, in the Local Standard of Rest (LSR) to similar values derived by \citet{tobin2009} from optical spectroscopy of a $\sim 2\deg$ FOV, for  $\sim$750 sources in both catalogs. Overall the agreement is good, with no noticeable zero-point offsets. The residual scatter may be due to observational errors in both measurements, as well as $v_r$ variations from stellar multiplicity.

One  feature in Figure \ref{figure:rv_comparison} is a modest population of datapoints { ($\sim 8\%$)} for which our $v_r$ are in line with the majority of the cluster members ($5~km~s^{-1}\leq v_r \leq 15~km~s^{-1}$), whereas the measurements from \citet{tobin2009} appear somewhat lower. As the right panel of \ref{figure:rv_comparison} shows the optical $v_r$ distribution is somewhat asymmetric, with the blue tail more prominent than the red one. On the other hand, the APOGEE $v_r$ distribution in the top panel of Figure \ref{figure:rv_comparison} does not present this skewness. This suggests that { a few} $v_r$  values measured by \citet{tobin2009} had been underestimated. { These few discrepant sources turn out to be nearly all in the low end of the temperature scale, with $T_{\rm eff}<4000$. One possibility is that previous results were affected by systematics in the M type spectral range, possibly due to inaccurate model spectra used to estimate $v_r$. Some of these sources have a moderately higher $A_V$ than average and, together with their low $T_{\rm eff}$ they occupy the faint end of the targeted sample. The observations of \citet{tobin2009} were conducted during an epoch with a negative heliocentric correction for Orion and in bright time. Thus low SNR and residual moonlight might have lead to underestimated velocities.}

A full analysis of the radial velocity distribution, in order to asses the kinematic status of the young stellar population throughout the Orion A cloud will be presented in a separate forthcoming paper (Da Rio et al., {\em in prep}).

\section{Stellar ages}
\label{section:ages}
The determination of stellar ages in young stellar clusters is exceptionally valuable for the study of star formation, as well as for planet formation, e.g., providing constraints to the circumstellar disk lifetimes. Spatial variations in ages along a molecular cloud are the footprints of how star formation progressed throughout space and time. { Similarly, the detection of a genuine spread in ages traces the duration of the star formation process, helping to constrain theoretical models and improve our understanding on the timescales for conversion of gas into stars \citep{elmegreen2000,hartmann2001,huff-staher2007,tan2006,longmore2014}.}

\begin{figure*}
\epsscale{0.8}
\plotone{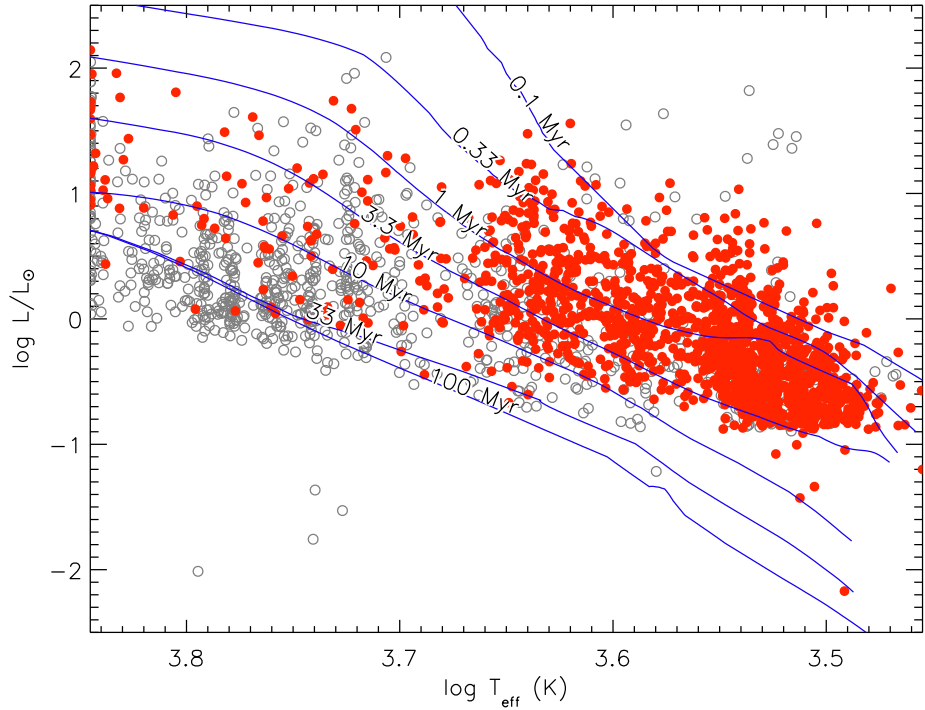}
\caption{HRD position derived for all our APOGEE targets. Red solid circles denote previously known members. The blue lines indicate isochrones from \citet{siess2000}, for ages of 0.1, 0.3, 1, 3, 10, 30, and 100~Myr. \label{figure:hrd}}
\end{figure*}
\begin{figure*}
\includegraphics[width=0.478\textwidth]{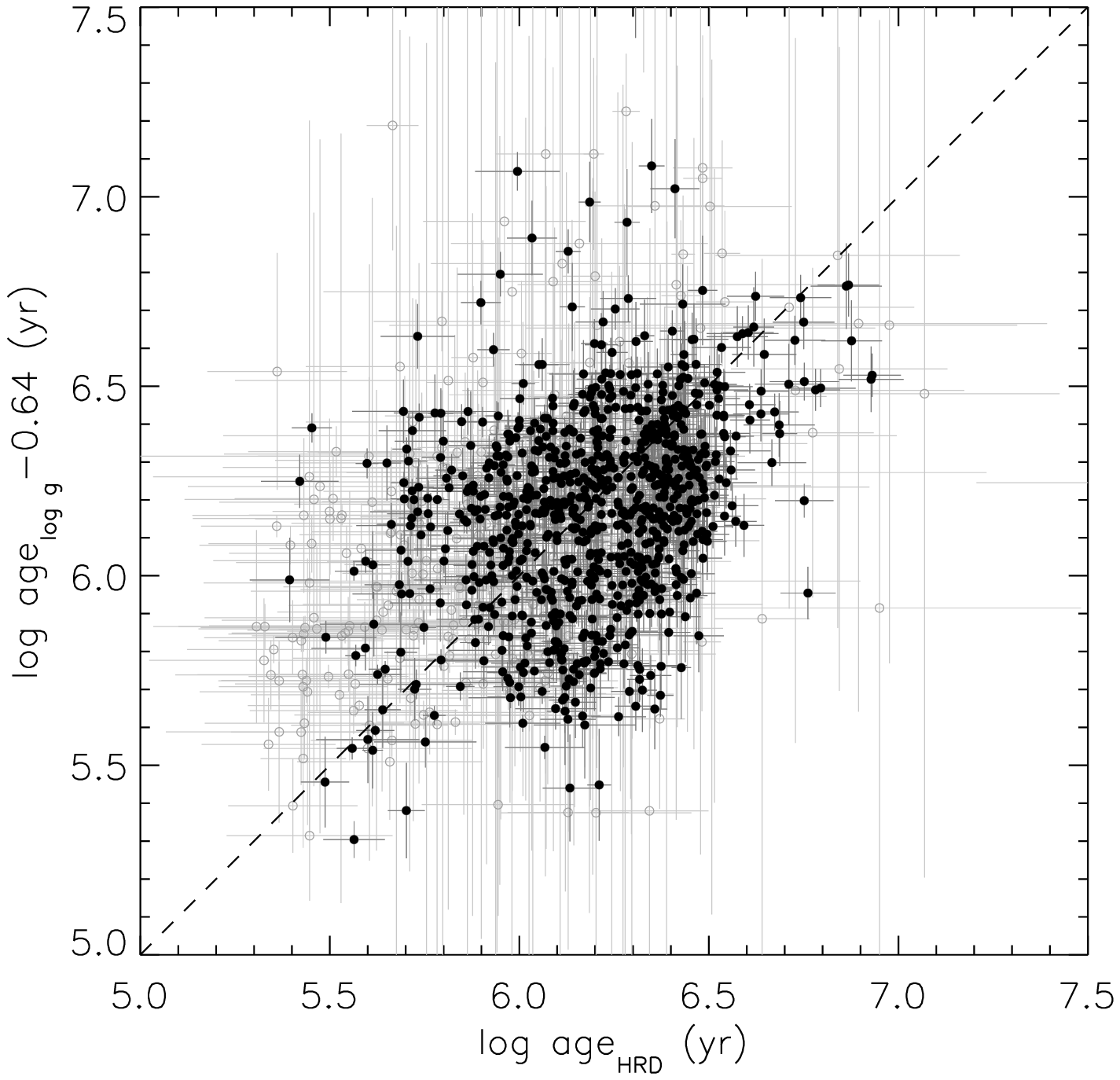}
\includegraphics[width=0.52\textwidth]{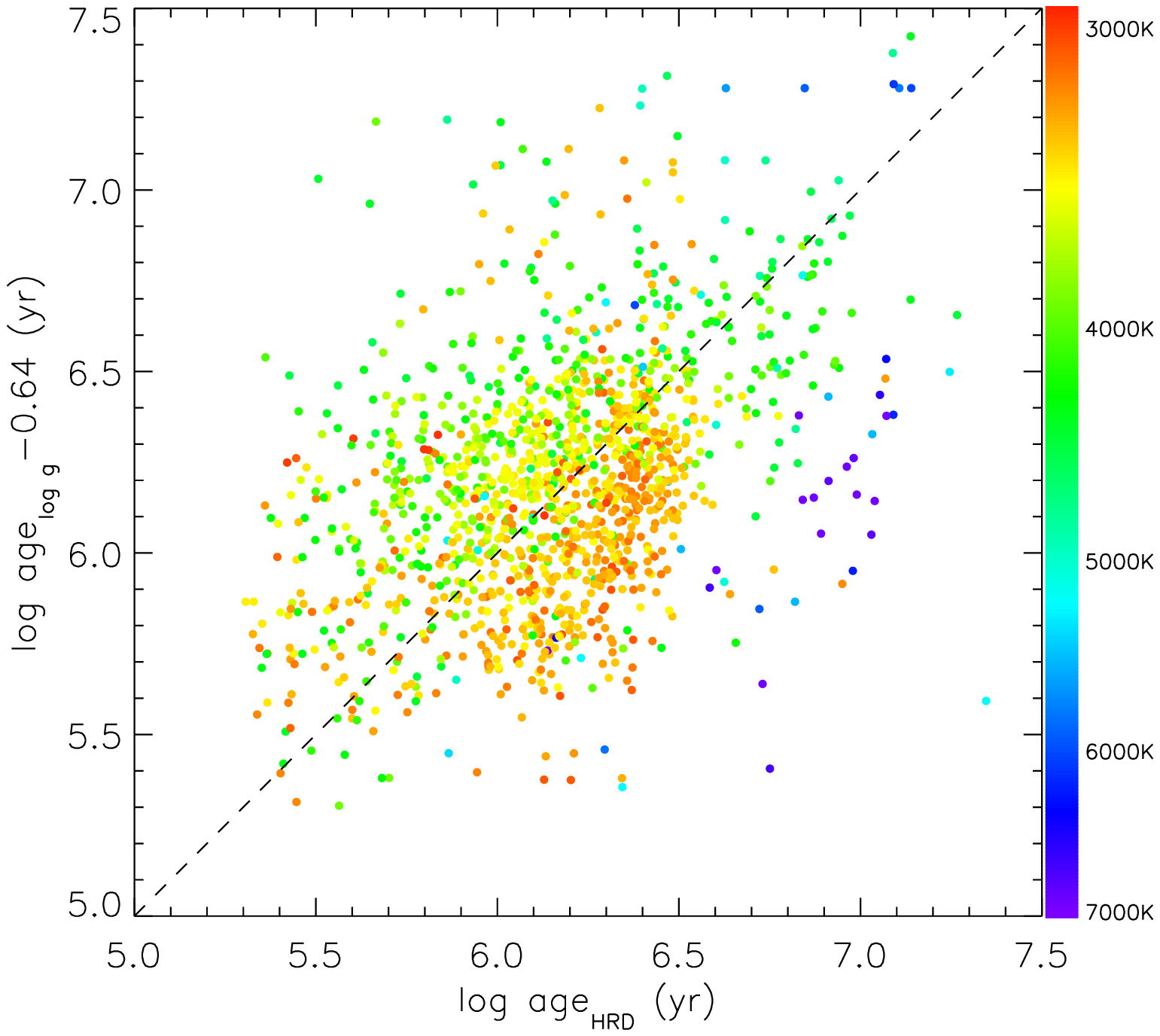}
\caption{{\em Left panel:} comparison between isochronal ages ($x$-axis) from the position of the sources in the HRD, and gravity-derived ages based on interpolated isochrones in the $\log g$--$T_{\rm eff}$ plane, the latter adjusted by a systematic shift of 0.64~dex. The grey circles indicate all members with $T_{\rm eff}<4200~K$, the black circles are those limited to small errors in both age estimates. {\em Right panel:} Same plot, color coded according to $T_{\rm eff}$. \label{figure:age_age}}
\end{figure*}

Yet, determining ages of young stars is known to be quite problematic \citep[see][for a review]{soderblom2014}. This is due to the difficulty in extracting their stellar parameters -- due to their complex spectra, variability and differential extinction \citep{dario2010a,dario2010b}, { unresolved multiplicity, binary evolution,} and the fact that PMS evolution may be affected by the protostellar accretion history \citep{baraffe2009,baraffe2012}. Although this last issue may not be dominant \citep{hosokawa2011}.
These limitations have put the reality of the age spreads, measured from the broadening of the PMS sequence of young clusters' HRDs, into question.
{ On one hand, young clusters are observed to be unembedded already at very young ages (~2 Myr, \citealp{portegies-zwart2010,longmore2014}), which limits the duration of ongoing star formation. On the other hand, very fast star formation scenarios are also problematic \citep{tan2006}.}
Several studies have been carried out to characterize the age distribution in the ONC: \citet{reggiani2011} showed that observational uncertainties in the construction of the HRD are minimal; \citet{jeffries2011}, from the lack of correlation between the fraction of stars bearing disks and their isochronal ages, concluded that the intrinsic width of the ONC's age distribution is less than half of the apparent width in logarithmic scale. Last, \citet{dario2014a} showed that a real age spread of at least 0.2 dex in the ONC (95\% of the population with ages between 1 and 6~Myr) must be invoked to justify the observed correlation between measured mass accretion rates and ages. To summarize, studies have shown that isochronal ages remain very uncertain, but this does not invalidate the existence of age spreads in young star forming regions including Orion. With all these caveats in mind, we attempt to study the age distribution in the Orion A cloud.

\begin{figure*}
\plottwo{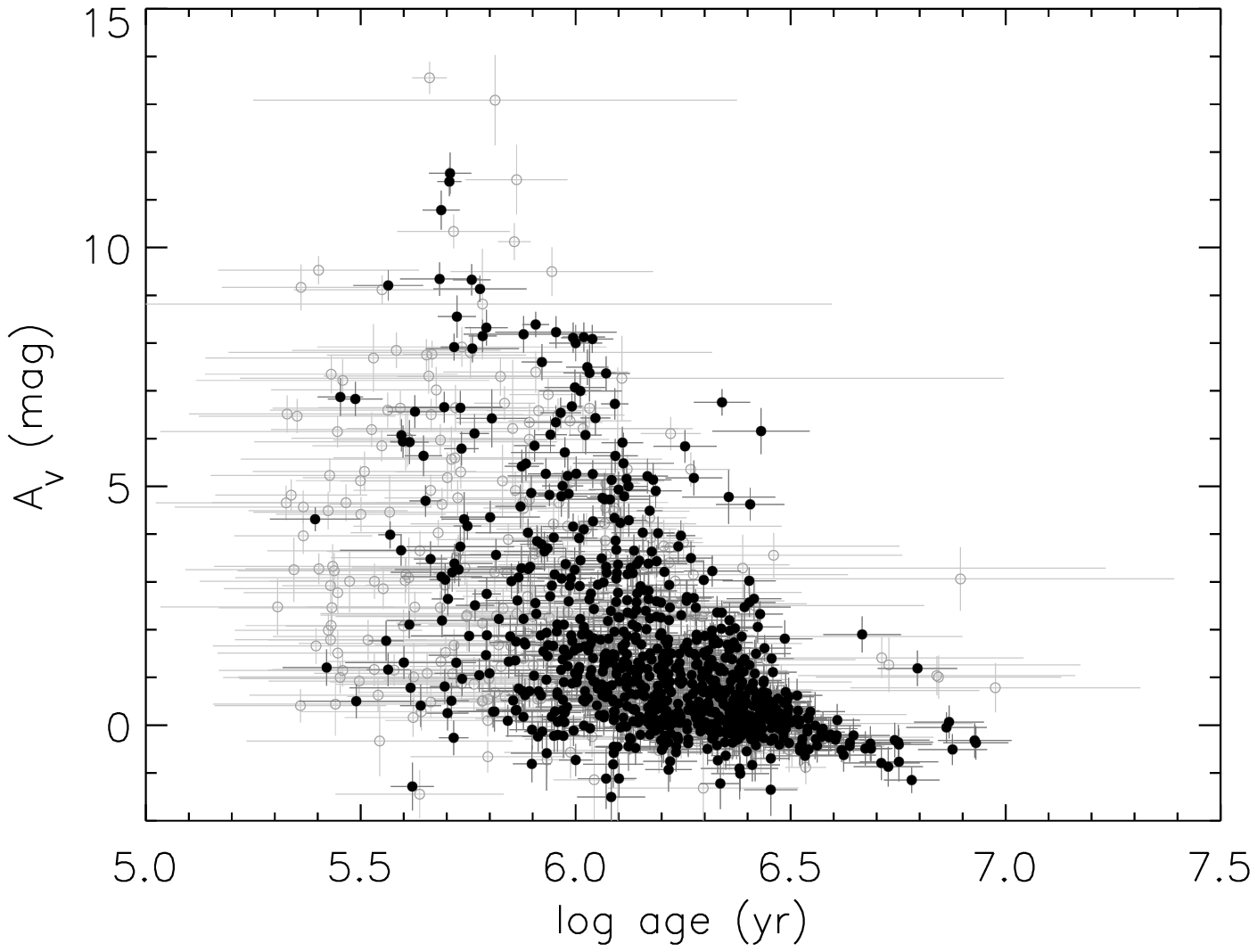}{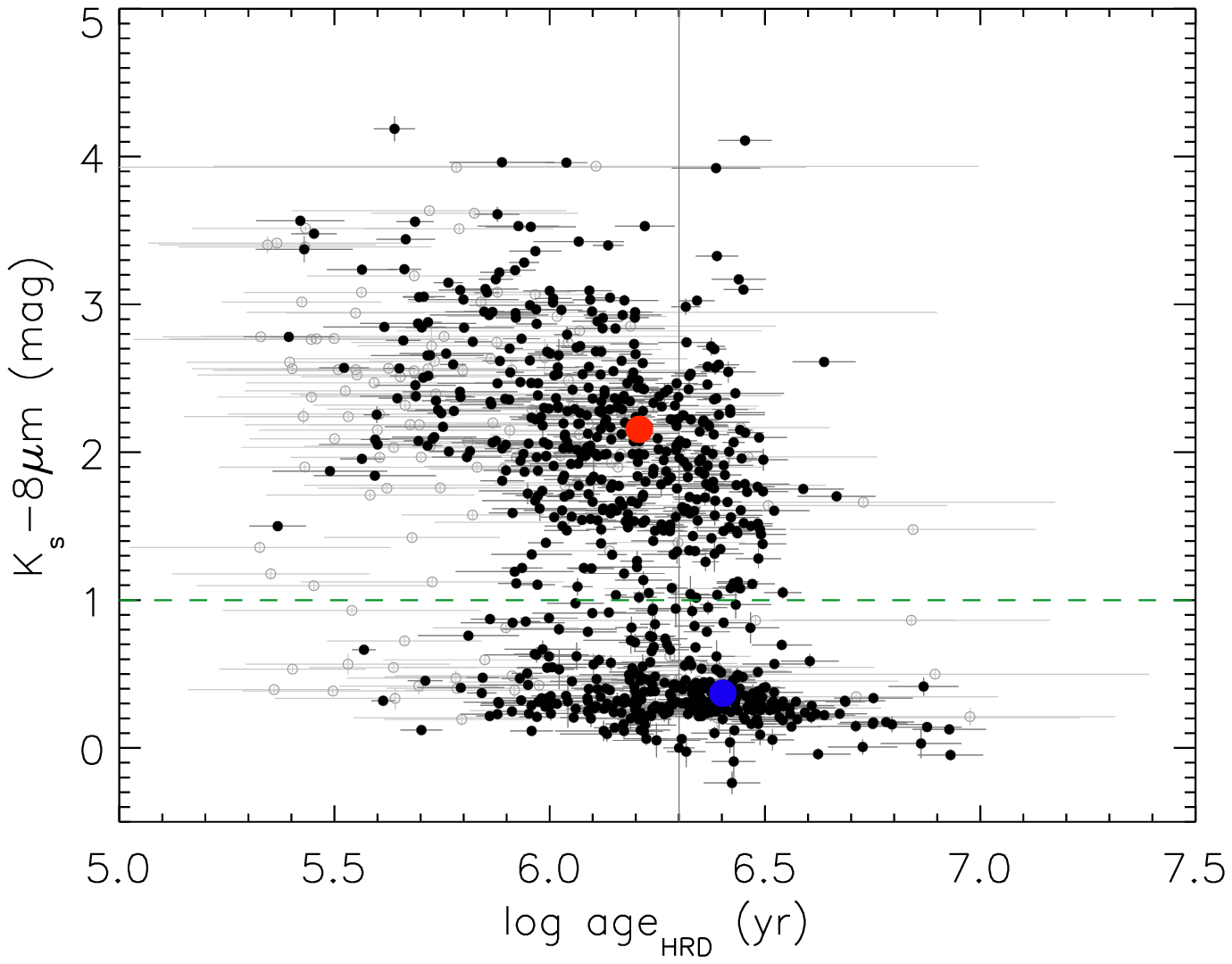}
\caption{{\em Left panel:} comparison between isochronal age and $A_V$. Open grey circles denote all members with $T_{\rm eff}<4200~K$, while filled black dots are those restricted to sources with small ($<0.15$dex) errors in ages. {\em Right panel:} same plot but comparing the age with the measured infrared SED slope. The two dots indicate the weighted mean of the subsamples, divided at $(K_s-8\mu m)=1$. \label{figure:age_av-irex}}
\end{figure*}

\subsection{ New evidence for a spread in stellar ages}
Figure \ref{figure:hrd} shows the HRD of the entire Orion A cloud survey: the luminosity has been derived from dereddened $J$ magnitudes, adopting our extinctions (Section \ref{section:stellar_parameters-av}) and corrected adopting the bolometric corrections from \citet{bell2014}. Isochrones from \citet{siess2000} are also shown. We assigned ages to each source adopting these models in two ways: interpolating the models into the HRD, and in the $\log g-T_{\rm eff}$ plane. The latter case, as shown in Figure \ref{figure:teff_logg}, is affected by systematic uncertainties. Although these systematics depend on $T_{\rm eff}$ and are likely nonlinear, we correct the surface gravity ages by a constant shift of 0.64~dex, which is the mean of the difference between all the ages according to the two methods. The comparison of the corrected ages from $\log g$ and $T_{\rm eff}$ is shown in Figure \ref{figure:age_age}, left panel, limited to known members and sources with $T_{\rm eff}<4200~K$, to avoid the poor accuracy in the stellar parameters at higher temperatures. Despite a scatter larger then the nominal average error bars, the correlation between the two is clear, with a measured Pearson correlation coefficient of $\rho\sim0.35$. Using a permutation test, we have established this correlation to be highly significant to a $\sim 11\sigma$ level. By comparing the scatter along the diagonal line in Figure \ref{figure:age_age} with that along its normal (indicative of anticorrelation), both weighted on the error bars of each point, we concluded that such correlation is indicative of an age spread of $0.16\pm0.01$~dex. This is smaller than the age spread estimated in the ONC, (0.2~dex, \citealt{dario2014a} or even 0.35~dex accounting only for HRD uncertainties \citealt{reggiani2011}), but it is only a lower limit. This is due to the fact systematic offsets in the estimated ages using independent method may vary across different regions of the stellar parameters space. This is evident from the right panel of Figure \ref{figure:age_age}, in which sources are color-coded according to their $T_{\rm eff}$: a clear gradient is evident, with age from $\log g$ being systematically underestimated for cool stars, and viceversa.
{ Overall, the measured apparent isochronal age spread is $\sim0.4$~dex, when measured locally given the age gradients along the Orion A filament (see Section \ref{section:ages-variations}). This is identical to that measured in the ONC from other studies \citep{hillenbrand1997,dario2012}.
We must highlight that this spread in ages identifies solely, strictly speaking, a spread in stellar radii, which may or may not indicate an age spread \citep{baraffe2012,jeffries2011}. This is because stellar evolutionary effects due to the past accretion history can potentially influence the radius of a PMS star for a given mass and age. As discussed in \citet{dario2014a}, however, the relation between mass accretion rates and isochronal ages implies a lower limit for the true age spread of $\sim0.2$~dex, corresponding, in the ONC, to the formation of $95\%$ of the stars within 5~Myr. Since the apparent spread in age we measure appears to be roughly constant throughout the entire cloud, we suggest that such relatively long duration of star formation occurred in the entire region. } We stress, however, that given our lack of knowledge on the 3D structure of the cloud part of this overall spread could be due to an age spread along the line of sight. This however is negligible at least in the ONC, the bulk of the unembedded population, likely distributed a few pc along the line of sight, shows evidence of a large spread in stellar ages.
In Paper I we also found evidence for a highly significant spread in stellar radii in IC348, based on APOGEE-derived surface gravities and isochronal ages.

We then utilize other indicators to better understand its nature. In Figure \ref{figure:age_av-irex}, left panel, we report the relation between $A_V$ and HRD ages, limiting to known members and highlighting sources with small uncertainty in ages with darker symbols. It appears that whereas young sources are distributed in a range of optical depths within the cloud, older members are basically found only at very low $A_V$, i.e. on the foreground end of the stellar distribution along the line of sight. This trend however is potentially suffering from two biases. First, if the extinction of a source had been highly overestimated, then so would its extinction-corrected $\log L$, and thus its age underestimated; vice versa for underestimated $A_V$.. Thus, errors in extinction tend to move sources in the $A_V$--age plane diagonally, in the same direction of the correlation observed. We exclude this bias to play a significant role here, because the correlation persists when we restrict to sources with minimal age (and thus $A_V$) errors (black circles in Figure \ref{figure:age_av-irex}, left panel) and also because of the good agreement between our APOGEE-derived $A_V$ and values from the literature (Figure \ref{figure:av_comparison}). A second bias could originate from incompleteness: older PMS stars are fainter than young ones, and the heavily  embedded ones will be likely too faint to be included in our sample. We tested this scenario isolating known members at intermediate temperatures ($T_{\rm eff}>4500$K), where incompleteness is quite low compared to the cold end of the population, and found that the correlation $A_V$--age remains identical. We therefore conclude that indeed older stars are slightly in foreground, whereas younger stars may be more embedded in the molecular material. This is not surprising considering that there is evidence the majority of the molecular material, at least in the ONC, is located behind the majority of the population along the line of sight \citep[see][]{dario2014b}. Within this material, dense cores are known to form stars today \citep{prisinzano2007}).

We also look for correlation between the isochronal ages and the circumstellar environment properties. In Figure \ref{figure:age_av-irex}, right panel, we plot the infrared slope ($K_s-8\mu m$), from 2MASS and SPITZER/IRAC respectively, with respect to stellar ages, again limiting to known members. We clearly detect sources with steeper slopes, indicative of earlier stages of stellar evolution, appear to have systematically younger ages derived from the HRD. { This fact also corroborates the fact that the spread in stellar radii is in fact indicative of the presence of genuine age spread, although the actual magnitude of this spread cannot be accurately constrained from this analysis.}

Using the ($K-8\mu m$)--age data, we can estimate the disk timescales. For simplicity we assume that the ratio of disk-bearing sources, in a given age interval of our target sample, is the ratio between the number of sources with $(K-8\mu m)>1$~mag over the total number. Limiting to known members with $T_{\rm eff}<4200$~K as in Figure \ref{figure:age_av-irex}, we determine an e-folding timescale for disk removal of $2.7\pm0.4$~Myr, consistent with previous findings \citep{fedele2010}. We also note that restricting to the ONC, the timescale increases, indicating weaker correlations between apparent age and disk properties, reaching up to 6~Myr within a radius of 0.5$\deg$ from the trapezium and 10~Myr within 0.25$\deg$. All these timescales however remain upper limits, as they are derived from the apparent measured ages. This is because when fitting the disk fractions versus ages additional scatter on the $x-$axis flattens the shape of the fitted function -- in this case an exponential -- because of regression dilution. { It should be also noted that in the densest core of the ONC, other dispersal mechanisms may affect the disk fraction, such as photoevaporation \citep{scally2001,robberto2002,adams2004} or dynamical truncation \citep{ovelar2012,rosotti2014}, although the latter mostly affects the external disk radius, and not the inner disk emission.}

\subsection{Spatial variations of stellar ages}
\label{section:ages-variations}
\begin{figure}
\epsscale{1.1}
\plotone{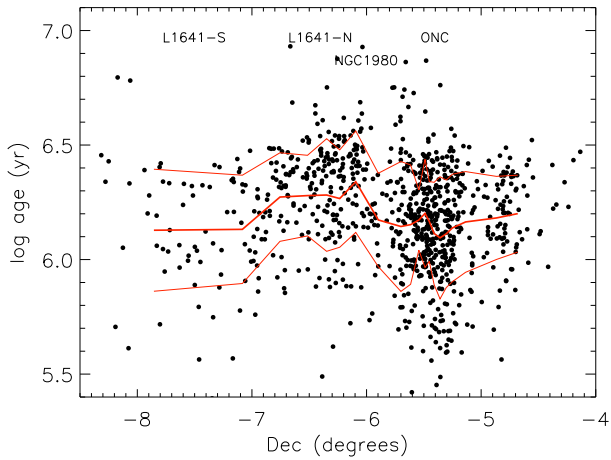}
\caption{Isochronal age versus declination, for all known members with $T_{\rm eff}<4200$~K. The thick line is the weighted average in different bins, whereas the thin lines identify the 1 sigma spread in ages, corrected for measurement errors. \label{figure:age_dec}}
\end{figure}
\begin{figure}
\plotone{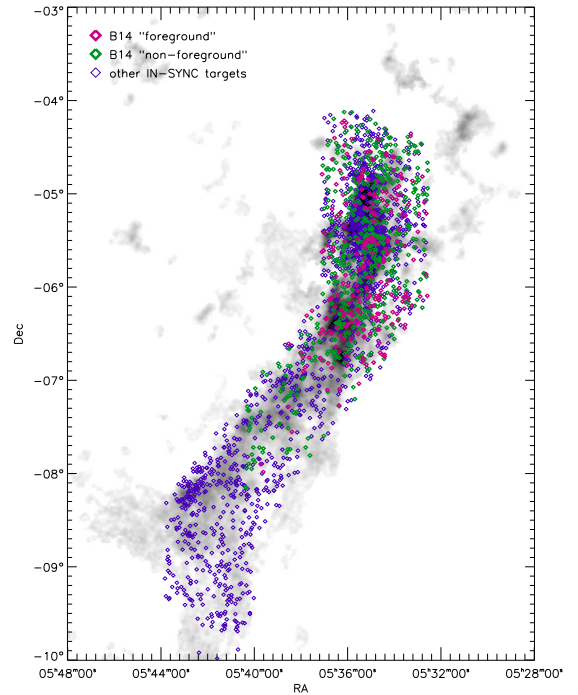}
\caption{Spatial distribution of our IN-SYNC sources over the \citet{nishimura2015} $^{13}$CO map. Sources flagged by \citet{bouy2014} as candidate of a separate ``foreground'' populations are marked as indicated by the legend. \label{figure:radec_bouy}}
\end{figure}
\begin{figure}
\plotone{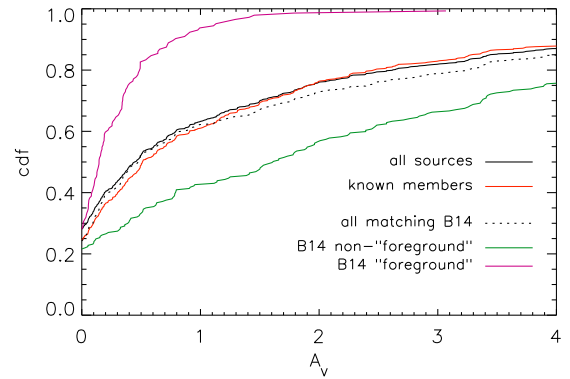}
\caption{Cumulative distributions for $A_V$ in different samples, as indicated by the legend. \label{figure:bouy_av_cdf}}
\end{figure}
\begin{figure*}
\plottwo{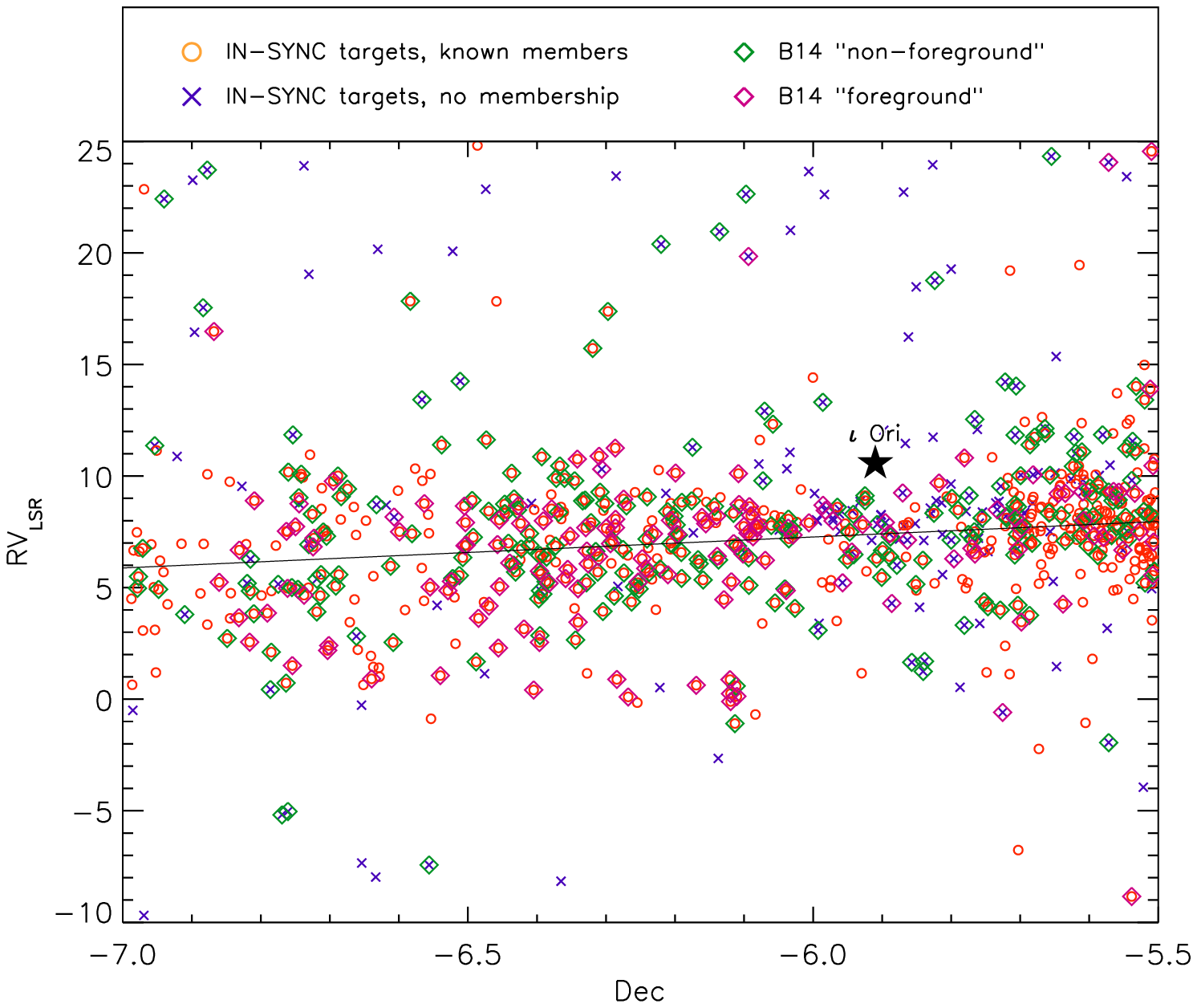}{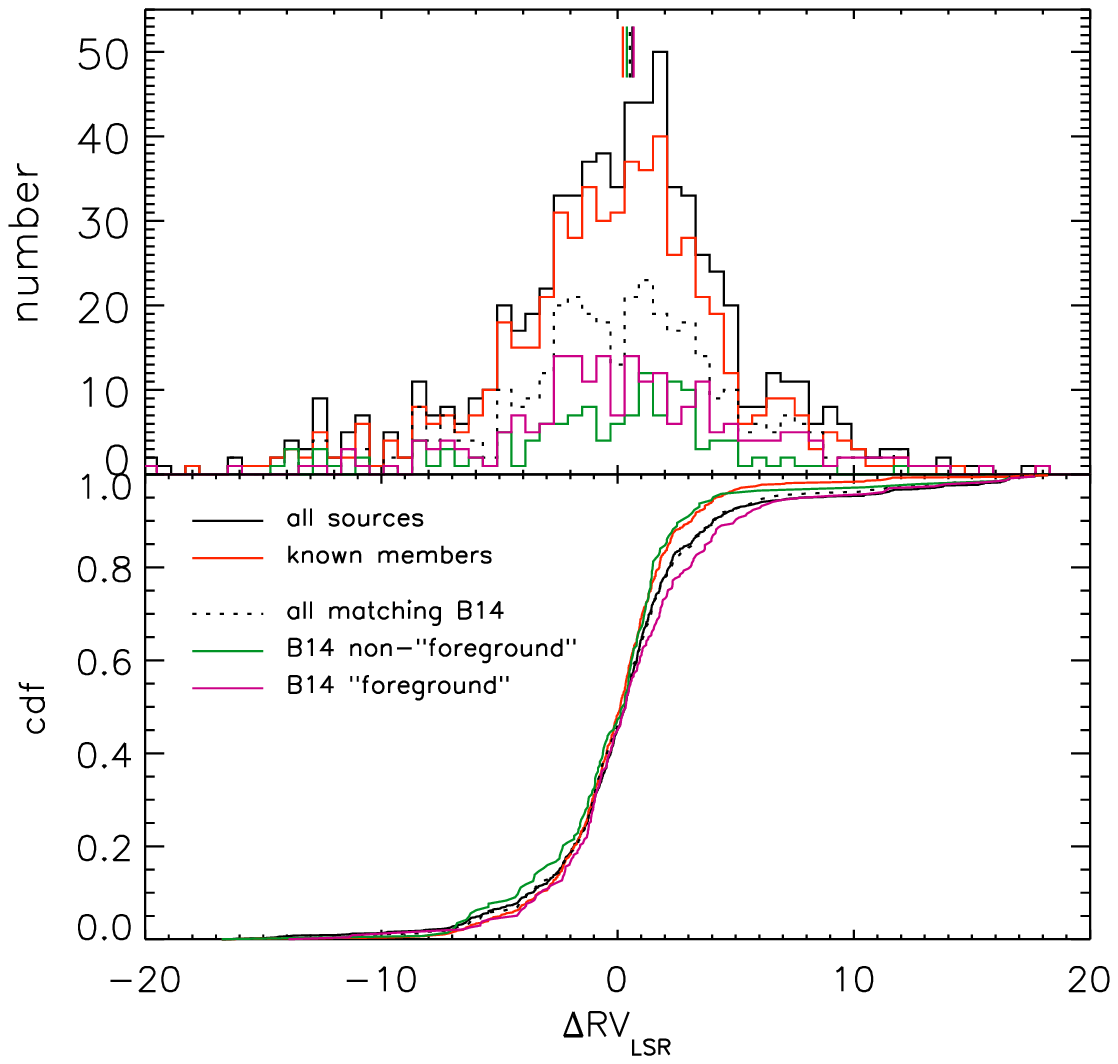}
\caption{{\em Left panel:} position velocity plot of all sources in the range $-7\deg<\delta<-5\deg30\arcmin$ around $\iota$ Ori and NGC1980. The solid line represents the linear fit which accounts to the large scale gradient in $v_r$. The star symbol denotes $\iota$ Ori. Stars in different samples are color coded as indicated in the legend. {\em Right panel:} $v_r$ distribution and cumulative distribution in the different samples. The vertical line segments indicate the weighted mean in each of them. \label{figure:rv_bouy}}
\end{figure*}
We look for systematic spatial variations of stellar ages along the Orion A cloud. Given the physical size of the cloud, about 40~pc assuming a distance of 414~pc, and the young age of the system there has been no time for a dynamical communication between the two ends of the filament, and thus no guarantee that the young population shares the same age throughout it. In Figure \ref{figure:age_dec} we plot the stellar ages as well as their local mean, as a function of declination. Whereas no large scale age gradients are detected, we find a statistically significant older age for the population at $\delta\simeq-6.2$, where the members are about 50\% older than, e.g., in the ONC ($\delta\simeq -5.5\deg$.)

\subsection{Is NGC1980 a foreground cluster?}
\label{section:ages-alves}
Following on the localized increase in stellar ages just mentioned, we note that a similar result was already presented in the literature. \citet{alves2012} suggested that the young population in the vicinity of $\iota$ Ori, identified with the association NGC1980, is in fact a separate population unrelated to the Orion A filament. Their claim was based on noticing that the majority of such sources are affected by very little extinction, thus possibly older (4-5~Myr old), and located in the foreground with respect to the rest of the region. Later \citet{bouy2014} used a multi-band photometric approach to isolate candidates of this population, confirmed its older age and estimated a distance to this system of $\sim 380$~pc, more than 30~pc closer to us than the rest of the young population.
We investigate these claims further based on our APOGEE data. { We cross match the catalogs from \citet{bouy2014} with our sample, and consider all sources present in both these samples. Among these, we further distinguish between stars attributed as ``foreground'' with a probability $p>70$\% from \citet{bouy2014} from the remaining which we refer to as ``non-foreground''} Figure \ref{figure:radec_bouy} shows the spatial distribution of these samples. The candidate population is indeed centered to the south of the ONC but is poorly concentrated and very spread out.

Figure \ref{figure:bouy_av_cdf} shows the cumulative distributions of the APOGEE derived $A_V$ values for sources in these different samples. All the sources in \citet{bouy2014} sample match closely the reddening distribution of our APOGEE survey. We confirm that the candidate foreground population is predominantly composed by low-$A_V$ sources, in line with the conclusion of their work. We also confirm these sources possess an older characteristic age: again, whereas the overall age, as derived both from our HRD and from the $\log g$--$T_{\rm eff}$ plane, of the entire \citet{bouy2014} catalog matches closely that of our survey, the candidate foreground population is $\sim$40\% older than the candidate non-foreground. { This is also understandable considering that the candidate foreground population is centered at declinations $-6.5\lesssim\delta\lesssim-6$, where the overall age of the Orion A cloud appears systematically older (see Figure \ref{figure:age_dec})}

We further analyze other indicators of age from the literature. We consider the the H$\alpha$ equivalent width ($EW$) estimates \citet{fang2013}, which were derived optical spectroscopy, for all sources matching both our APOGEE sample and that of \citet{bouy2014}, and restrict the sample in a declination range $-7\deg<\delta<-5\deg30\arcmin$, in order to bracket more closely the area around NGC1980. We separate classical T-Tauri sources (CTTS) as those showing $EW>20\AA$ from the remaining weak-line T-Tauri stars (WTTS). Table \ref{table:bouy_TTS} reports the numbers in the different stellar samples. The sample from \citet{bouy2014} matches all sources in our APOGEE survey within the declination range we considered. The candidate ``foreground'' population, however, shows a 6-fold excess in the WTTS/CTTS ratio compared to the remaining ``non-foreground'', although this difference is statistically significant only at $\sim1.8\sigma$ due to the low numbers of sources included. This finding further supports the older age of the candidate ``foreground'' sources. Similarly, we consider the disk classification from \citet{fang2013}, which, based on infrared SEDs, separates sources bearing thick disks, thin disks, transition disks or diskless members. Table \ref{table:bouy_disks} reports the numbers in each category for the different stellar samples. Again, the candidate ``foreground'' populations shows a larger fractional excess of transition disks and diskless stars compared to the ``non-foreground'' sample, confirming a more evolved population.

\begin{table}
\caption{T-Tauri star classification within $-7<\delta<-5\deg 30\arcmin$\label{table:bouy_TTS}}
\begin{tabular}{l|r|r|r|r|}
  & CCTS & WTTS & Unknown & WTTS/CTTS \\[1ex] \hline
all sources            & 44 & 122 & 596 & 2.8$\pm$0.5 \\[1ex]
all in B14             & 26 & 80  & 275 & 3.1$\pm$0.7 \\[1ex]
B14 ``foreground''     & 5  & 46  & 93  & 9.2$\pm$4.3 \\[1ex]
B14 ``non-foreg.''     & 21 & 34  & 182 & 1.6$\pm$0.4 \\[1ex] \hline
\end{tabular}
\end{table}

\begin{table}
\caption{Disk properties within $-7<\delta<-5\deg 30\arcmin$ \label{table:bouy_disks}}
\begin{tabular}{l|r|r|r|r|r|}
  & Thick & Thin & Trans. & No Disk & unknown \\[1ex] \hline
all sources            &   45    &      3    &     14   &      105  &      595 \\[1ex]
all in B14             &   28    &      1    &     11   &      67  &      247 \\[1ex]
B14 ``foreground''     &    6    &      0    &      7   &      38  &       93\\ [1ex]
B14 ``non-foreg''      &   22    &      1    &      4   &      29  &      181 \\[1ex] \hline
\end{tabular}
\end{table}

We have confirmed that the proposed ``foreground'' population isolated by \citet{bouy2014} is indeed both older and affected by less extinction. We now analyze its kinematic properties. Figure \ref{figure:rv_bouy}, left panel, reports the position-$v_r$ diagram for the declination range around NGC1980. For APOGEE targets observed in more than one epoch we have considered a weighted mean of the individual $v_r$ values. Since the entire Orion A cloud presents a large scale gradient in $v_r$ from north to south, observed both in the gas kinematics \citep{bally1987,nishimura2015} and in the stellar motions (\citealt{tobin2009}, Da Rio et al. (2015b, {\em in prep}), we isolated this gradient using a linear regression and extracted the residuals $\Delta v_r$ for each source with respect to the mean velocity at each declination. Figure \ref{figure:rv_bouy}, right panel, reports both the $v_r$ distribution of the residuals and the cumulative distribution. We caution the reader that such distributions are not corrected for $v_r$ uncertainties and scatter from multiplicity; a full kinematic analysis throughout the region is to be presented in Da Rio et al. (2015b, {\em in prep}), however this is sufficient for a meaningful comparison between the different samples. We excluded from the analysis stars with $v_r$ residuals larger than 20~km~s$^{-1}$ as outliers. Although the mean $v_r$ show some small offsets between the different samples, with the candidate ``foreground population'' slightly blue shifted compared to the ``non foreground'' sample, the distributions appear very similar. A K-S test on the cumulative distributions could not rejects  the hypothesis that the $v_r$ distribution of the foreground and non-foreground populations are identical, with high significance ($p<10^{-6}$).

\begin{figure*}
\epsscale{1.15}
\plotone{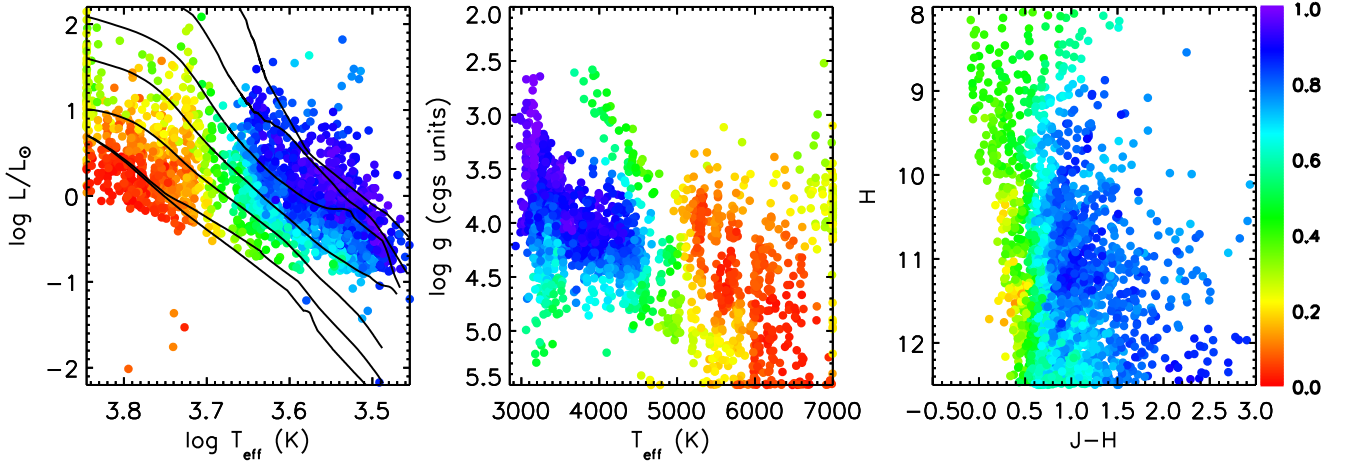}
\caption{HRD, $T_{\rm eff}$--$\log g$ and CMD planes, with sources color coded to the local fraction of known members over the local total of targets, as indicated in the legend. \label{figure:membership_planes}}
\end{figure*}
\begin{figure}
\plotone{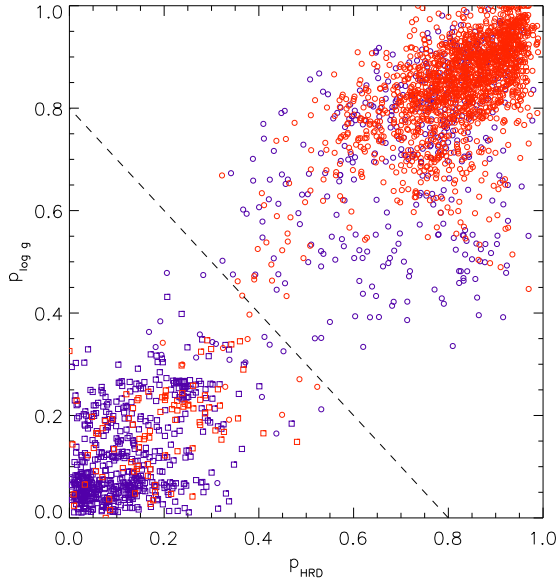}
\caption{The comparison between $p_{HRD}$ and $p_{\log g}$ shows a clear correlation, indicating that member-dominated regions of the HRD  correspond to member-dominated regions of the $T_{\rm eff}$--$\log g$. Red circles indicate known members from the literature, blue circles sources with no confirmed membership. The dashed line indicated the threshold $p_{HRD}+p_{\log g}=0.8$ we imposed, only for sources with $T_{\rm eff}<5000$, above which a star must lie as one of the conditions to be flagged as new candidate member. \label{figure:mem_hrd_logg_known}}
\end{figure}
\begin{figure}
\plotone{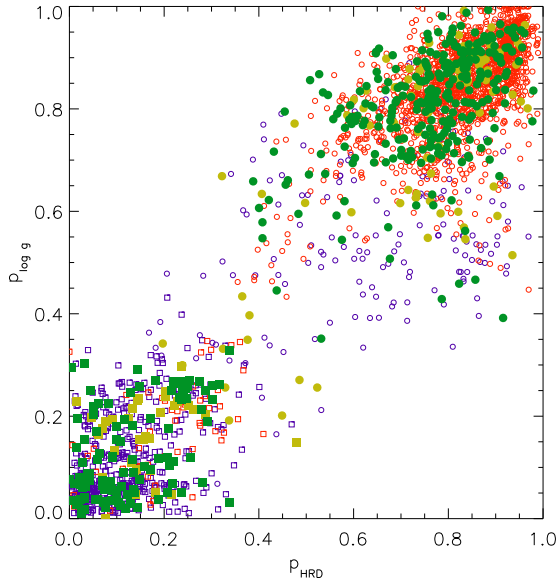}
\caption{Same as Figure \ref{figure:mem_hrd_logg_known}, but highlighting sources flagged as new candidate members from our technique (green dots) as well as known members that were missed (yellow dots). Red circles denote known members that were also recovered, whereas blue circles sources with no previous membership estimates that are now identified as candidate non-members. \label{figure:mem_hrd_logg_new}}
\end{figure}
\label{section:membership}
\begin{figure*}
\epsscale{1.15}
\plotone{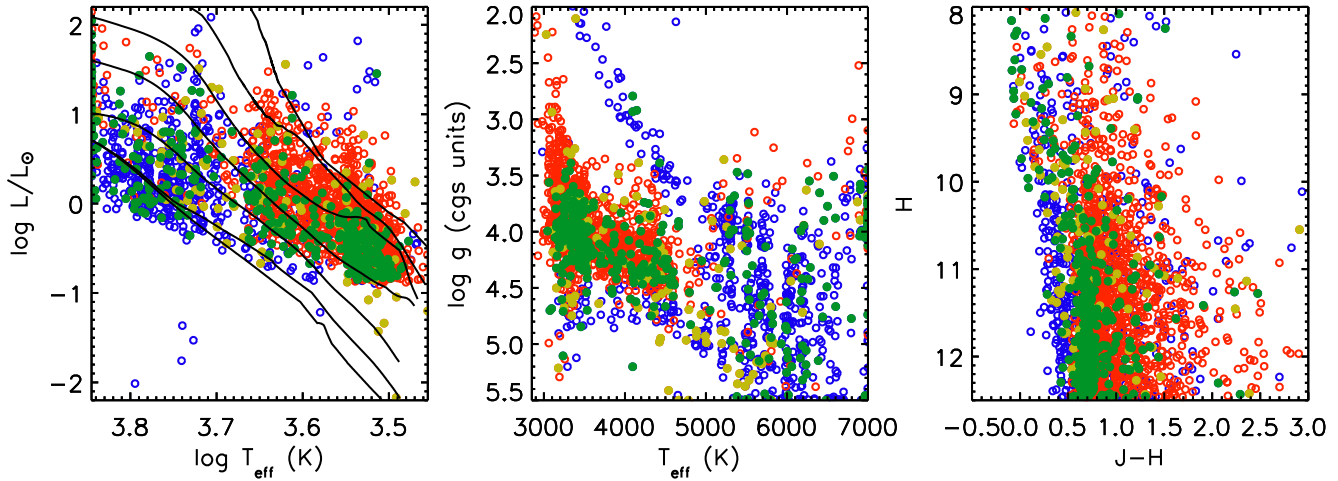}
\caption{{ The 3 planes as in Figure \ref{figure:membership_planes}, with sources color-coded as Figure \ref{figure:mem_hrd_logg_new}.} \label{figure:membership_new}}
\end{figure*}
\begin{figure}
\epsscale{1.15}
\plotone{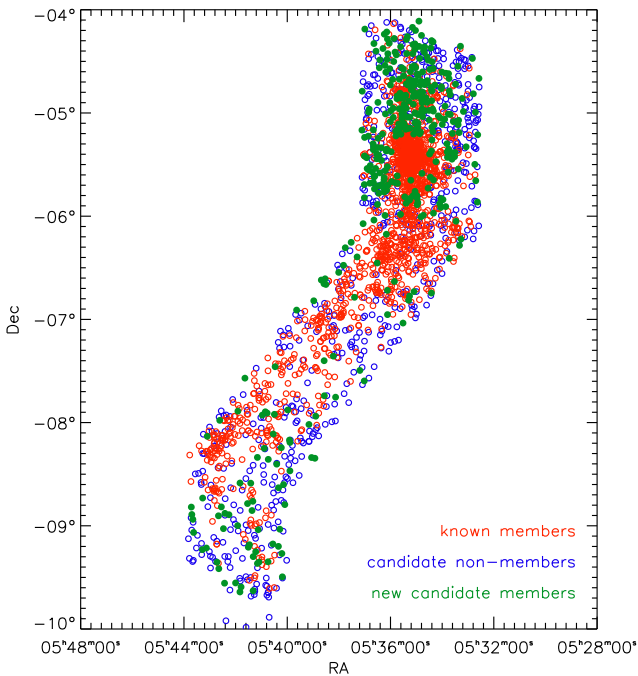}
\caption{{ Spatial distribution of sources; open red circles represents previously known members, filled green circles new candidate members from this work, open blue circles sources with no literature membership that have not been flagged as candidate members from this work.} \label{figure:mem_radec}}
\end{figure}

\begin{table*}
\centering
\caption{Fitted stellar parameters. \label{table:fitted_parameters}}
\begin{tabular}{rrcrrrrrrr}
\hline
\hline
\multicolumn{1}{c}{R.A}. & \multicolumn{1}{c}{dec} &  & \multicolumn{1}{c}{$T_{\rm eff}$} & \multicolumn{1}{c}{$\log g$} & \multicolumn{1}{c}{$v\sin i$} & \multicolumn{1}{c}{$J$} & \multicolumn{1}{c}{$H$} & \multicolumn{1}{c}{$A_V$}\vspace{-0.2cm} \\
\vspace{-0.2cm} &  & \multicolumn{1}{c}{epochs} & & & & & & \\
\multicolumn{1}{c}{(J2000)}\vspace{-0.2cm} & \multicolumn{1}{c}{(J2000)} & & \multicolumn{1}{c}{K} & \multicolumn{1}{c}{cgs} & \multicolumn{1}{c}{km s$^{-1}$} & \multicolumn{1}{c}{mag} & \multicolumn{1}{c}{mag} & \multicolumn{1}{c}{mag}\vspace{0.2cm}  \\
\hline
5$^h$32$^m$33.99$^s$ & -4$\deg$44$\arcmin$55.65$\arcsec$ & 1 & 5505$\pm$29 & 3.98$\pm$0.08 & 4.48$\pm$1.47 & 13.11$\pm$0.02 & 12.13$\pm$0.02 & 5.07$\pm$0.26 \\
5$^h$32$^m$34.29$^s$ & -4$\deg$56$\arcmin$25.47$\arcsec$ & 1 & 4381$\pm$25 & 3.53$\pm$0.07 & 4.72$\pm$1.29 & 12.54$\pm$0.02 & 11.65$\pm$0.02 & 2.12$\pm$0.26 \\
5$^h$32$^m$34.57$^s$ & -4$\deg$39$\arcmin$54.63$\arcsec$ & 1 & 5982$\pm$87 & 4.85$\pm$0.14 & 6.63$\pm$1.77 & 12.98$\pm$0.03 & 12.31$\pm$0.03 & 3.33$\pm$0.31 \\
5$^h$32$^m$35.05$^s$ & -6$\deg$5$\arcmin$37.31$\arcsec$ & 1 & 3322$\pm$26 & 4.41$\pm$0.11 & 0.74$\pm$2.53 & 11.38$\pm$0.02 & 10.82$\pm$0.02 & -0.72$\pm$0.23 \\
5$^h$32$^m$36.16$^s$ & -5$\deg$0$\arcmin$41.98$\arcsec$ & 4 & 3182$\pm$24 & 4.91$\pm$0.14 & 9.65$\pm$1.45 & 12.72$\pm$0.02 & 12.10$\pm$0.02 & -0.13$\pm$0.22 \\
5$^h$32$^m$37.16$^s$ & -5$\deg$53$\arcmin$21.53$\arcsec$ & 1 & 5751$\pm$94 & 5.50$\pm$0.02 & 0.68$\pm$3.53 & 12.38$\pm$0.02 & 11.84$\pm$0.02 & 1.84$\pm$0.26 \\
5$^h$32$^m$37.45$^s$ & -5$\deg$4$\arcmin$2.33$\arcsec$ & 1 & 5530$\pm$16 & 4.00$\pm$0.04 & 3.32$\pm$0.91 & 12.12$\pm$0.02 & 11.44$\pm$0.02 & 2.66$\pm$0.24 \\
5$^h$32$^m$37.49$^s$ & -5$\deg$12$\arcmin$41.28$\arcsec$ & 1 & 3595$\pm$11 & 4.21$\pm$0.03 & 12.02$\pm$0.43 & 12.44$\pm$0.03 & 11.69$\pm$0.02 & 0.17$\pm$0.23 \\
5$^h$32$^m$38.05$^s$ & -5$\deg$51$\arcmin$48.41$\arcsec$ & 1 & 3459$\pm$11 & 3.93$\pm$0.04 & 11.46$\pm$0.45 & 11.66$\pm$0.02 & 10.95$\pm$0.02 & 0.13$\pm$0.23 \\
5$^h$32$^m$38.46$^s$ & -5$\deg$31$\arcmin$18.09$\arcsec$ & 1 & 5590$\pm$29 & 4.79$\pm$0.05 & 6.24$\pm$0.60 & 10.35$\pm$0.02 & 10.07$\pm$0.02 & -0.48$\pm$0.22 \\
... & ... & ... & ... & ... & ... & ... & ... & ...  \\
\hline
\hline
\end{tabular}
\begin{tablenotes}
\item (This table is available in its entirety in a machine-readable form in the online journal. A portion is shown here for guidance regarding its form and content.)
\end{tablenotes}
\end{table*}

\begin{table*}
\centering
\caption{Additional quantities. \label{table:additional_parameters}}
\begin{tabular}{rrr|rrr|r}
\hline
\hline
\multicolumn{3}{c|}{} & \multicolumn{3}{c}{literature memberships} & \multicolumn{1}{|c}{new} \\
\multicolumn{3}{c|}{}\vspace{-0.2cm} & \multicolumn{3}{c}{ } & \multicolumn{1}{|c}{} \\
\multicolumn{1}{c}{$\log L$}\vspace{-0.2cm} & \multicolumn{1}{c}{$\log age$} & \multicolumn{1}{c|}{$M$} & \multicolumn{1}{c}{optical} & \multicolumn{1}{c}{IR} & \multicolumn{1}{c}{} & \multicolumn{1}{|c}{candidate} \\
\vspace{-0.2cm} &  & \multicolumn{1}{c|}{} &   &   & \multicolumn{1}{c}{X-ray$^c$}  & \multicolumn{1}{|c}{} \\
\multicolumn{1}{c}{($L_{\odot}$)}\vspace{-0.2cm} & \multicolumn{1}{c}{(yr)} & \multicolumn{1}{c|}{M$_\odot$} & \multicolumn{1}{c}{spectr.$^a$}  & \multicolumn{1}{c}{excess$^b$}  &   & \multicolumn{1}{|c}{members} \vspace{0.2cm}  \\
\hline
0.115$\pm$0.037 & 7.39$\pm$0.04 & 1.12$\pm$0.03 & \xmark & \xmark & \xmark & \xmark \\
-0.194$\pm$0.038 & 6.93$\pm$0.06 & 1.10$\pm$0.02 & \xmark & \xmark & \xmark & \xmark \\
-0.012$\pm$0.045 & 8.02$\pm$0.02 & 1.15$\pm$0.02 & \xmark & \xmark & \xmark & $\checkmark$ \\
-0.245$\pm$0.032 & 5.99$\pm$0.08 & 0.29$\pm$0.01 & \xmark & \xmark & \xmark & \xmark \\
-0.716$\pm$0.032 & 6.37$\pm$0.02 & 0.22$\pm$0.01 & \xmark & \xmark & \xmark & \xmark \\
-0.000$\pm$0.039 & 7.97$\pm$0.20 & 1.11$\pm$0.02 & \xmark & \xmark & \xmark & \xmark \\
0.186$\pm$0.031 & 7.34$\pm$0.02 & 1.16$\pm$0.02 & \xmark & \xmark & \xmark & $\checkmark$ \\
-0.502$\pm$0.033 & 6.37$\pm$0.04 & 0.40$\pm$0.01 & \xmark & \xmark & \xmark & $\checkmark$ \\
-0.216$\pm$0.031 & 6.07$\pm$0.02 & 0.34$\pm$0.00 & \xmark & \xmark & \xmark & \xmark \\
0.477$\pm$0.033 & 7.09$\pm$0.04 & 1.42$\pm$0.04 & \xmark & \xmark & \xmark & \xmark \\
... & ... & ... & ... & ... & ... & ...  \\
\hline
\hline
\end{tabular}
\begin{tablenotes}
\item (This table is available in its entirety in a machine-readable form in the online journal. A portion is shown here for guidance regarding its form and content.)
\item $^a$ -- membership from optical spectroscopy \citep{hillenbrand1997,dario2012,hsu2012,hsu2013,fang2009,fang2013}.
\item $^b$ --  \citep{getman2005,pillitteri2013}.
\item $^c$ -- \citep{megeath2012}.
\end{tablenotes}
\end{table*}

{
Given that the kinematic properties of the candidate foreground cluster proposed by \citet{alves2012} and \citet{bouy2014} are indistinguishable from those of the rest of the population at the same position in the sky, we find it unlikely that this population is a separate entity from the rest of the Orion A young population, and located tens of parsecs closer along the line of sight. Instead, we find it more likely that such population simply represents the older tail of the age distribution around $\delta\sim6\deg$, in the context of a long duration star formation event.
Our proposed scenario mainly leverages on the kinematic evidence. The entire Orion A cloud shows a $v_r$ gradient of 10~km~s$^-1$, with the north end receding and the south end approaching. If the foreground population were separated by several tens of pc, which is about the size of the Orion A cloud itself, it would be very coincidental for it to share the identical mean $v_r$ as the portion of the younger Orion A population behind it. This would make the specific direction of the line of sight somewhat special: on one hand because the foreground population would be perfectly aligned in projection against the Orion A filament, on the other hand if an observed had to look at it from a slightly different angle, its mean $v_r$ (and possibly the $v_r$ dispersion) would not match that of another portion of the young population seen in its background.
We do not exclude that the older populations remain somewhat in the foreground of the Orion A cloud, but it seems likely that any separation between the two should be minimal, and the hypothesis that it originates from an independent star formation event would be unfavourable.
Also, the low $A_V$ of the older population may not necessarily imply a closer distance to us, but simply a lower density of molecular material surrounding it, which is expected for a more evolved young population.
Thus, we speculate that start formation in the Orion A cloud -- which developed over several Myr in each location -- peaked earlier in the region around NGC1980, and later in the remaining areas.

}

\section{Memberships}
\label{section:membership}

Based on all data and parameters we collected from our APOGEE spectra, we can improve membership determination for all sources with no membership derived in the literature. Specifically, we consider the location of all the sources in a number of planes: a) the CMD $H$--($J-H$), b) the HRD, c) the $\log g$--$T_{\rm eff}$ plane, and d) the position-velocity plane $v_r$--$\delta$. For the first 3 cases, it is evident from Figure \ref{figure:cmds}, \ref{figure:hrd} and \ref{figure:teff_logg} that known members tend to occupy well confined regions of these planes. In the CMD they are confided to a locus that is redder than a young PMS isochrone, whereas MS contaminants have bluer colors. In the HRD and $T_{\rm eff}$--$\log g$ planes at low $T_{\rm eff}$ ($<5000$~K) known members occupy a tighter sequence compared to the rest of the unclassified sources. At hotter temperatures $T_{\rm eff}$, $\log L$ and $\log g$ have little to no diagnostic power to isolate members, but as anticipated the large number of sources in this range suggests that the vast majority of them are non-members.

For every star, in each of these 3 planes we consider the 50 closest sources to the considered star, where the distance is measured in units of measurement errors in the two parameters of the plane. We then count among the 50 sources the fraction that have been previously identified as members, and consider this value, regardless of our knowledge of the membership from the literature, a basic indicator the membership probability for the selected star in that given plane. We refer to these three quantities as $p_{HRD}$, $p_{\log g}$ and $p_{CMD}$. Figure \ref{figure:membership_planes} shows the three planes, with sources color-coded according to the value of $p$ for each specific plane. Figure \ref{figure:mem_hrd_logg_known} compares $p_{HRD}$, $p_{\log g}$. A clear correlation is evident and two populations appear separated at high and low membership probability. { Moreover, known members from the literature show high values of $p$, whereas sources without a literature membership are mostly distributed in the low $p$ end of this diagram.}
 A few sources with no previous memberships still populate the top-right corner, and among these some could be new actual members. On the other hand some known members show very low membership probability from these 2 planes, and these are all intermediate $T_{\rm eff}$ stars. A similar comparison using $p_{CMD}$ estimates shows much less correlation with the other 2, due to the  ability of this plane to { only} identify low $A_V$ field contaminants as systematically bluer then members in $J-H$.

Last we consider the radial velocity distribution of all the sources, compared to the average one of the known members. Given the large scale gradients of the mean velocity along the filament, which is also observed in the gas \citep{bally1987,nishimura2015}, we construct a 2 dimensional map of the average $v_r$ as well as the velocity dispersion considering the known members. To this end, for each point in RA and Dec of our field of view, we consider a circular aperture of 20\arcmin radius. If this contains at least 30 members, we measure their $\overline{v_r}$ and $\sigma_r$, weighting on the errors and excluding outliers with a sigma clipping algorithm ($\sigma>3)$. Note that for this application the local $\sigma_r$ was not corrected for unresolved multiplicity. A full analysis of the stellar velocity dispersion throughout the Orion A cloud is to be presented in a forthcoming paper. For each star, we then computed the velocity offsets $\Delta v_r$, in units of standard deviations, from the local $\overline{v_r}(\alpha,\delta)$, where the standard deviation comprises the measurement error in $\delta v_r$, the local $\sigma_r$, as well as the scatter in the individual measurements of $v_r$ for the star, if observed in more than one epoch and larger $\delta v_r$, which is the case of, e.g., $v_r$ variations from binarity.

{
From all $p_{HRD}$, $p_{\log g}$ and $p_{CMD}$ and $\Delta v_r$ for all our targets, we attempt to estimate memberships for sources with no membership estimates from the literature. A rigorous approach that utilizes these four parameters as formal probabilities cannot be applied, since none of them is a probability, but they represent arbitrary parameters with variable diagnostic ability to trace membership in the parameter space. For instance, $p_{HRD}$, $p_{\log g}$ and $p_{CMD}$ are not fully independent from each other, since positions in the HRD and in the CMD are related for less than the differential extinction. Also, $\log g$ and $\log L$ are correlated for any given temperature. As anticipated, $p_{HRD}$, $p_{\log g}$ have little diagnostic power for the hotter sources, because this $T_{\rm eff}$ range is dominated by contaminants. Therefore we  look for for the optimal, while arbitrary, combination of constraints that reasonably maximize the number of new candidate members we identify while safely reducing the number of false candidates. A good combination of criteria must recover, as candidate members, the largest number of known members from the literature, while also maximising the number of new candidates previously uncategorized.

For for sources with $T_{\rm eff}>5000$~K we simply impose that the $\Delta v_r <2.5$, i.e., within the 99\% percentile of the local mean stellar velocity, and also impose $p_{CMD}>0.3$, to exclude MS sources bluer than the PMS sequence of the young population. For the colder end of the population ($T_{\rm eff}<5000$~K), where both the HRD and the $T_{\rm eff}$--$\log g$ planes allow to separate well PMS stars from contaminants, we impose, in addition on the constraints for hotter sources, also $p_{HRD}+p_{\log g}>0.8$, i.e., above the dashed line in Figure \ref{figure:mem_hrd_logg_known}.
With these constraints we were able to recover 95\% of the known members. The remaining 120 known members were not selected primarily due to discrepant $v_r$ (101 sources), likely due to binarity.
In addition, our membership criterion selected 383 new sources as candidate members. Figures \ref{figure:mem_hrd_logg_new} and \ref{figure:membership_new} show the new candidate members in the same planes explored before.

As evident also from figures \ref{figure:membership_new}, the new identified candidate members are mostly in the cool end of the $T_{\rm eff}$ scale, with 75\% of them with $T_{\rm eff}<5000$~K. Moreover, nearly all of them appear to be class III sources, with 95\% of them showing an IR slope ($K-8\mu m$)$<1$. This is visible in all panels of \ref{figure:membership_new}: new cool candidates tend to show lower luminosities and higher $\log g$ compared to the mean of the known members, while their location in these panels is in agreement with the overall population. The 25\% of the new candidate members with $T_{\rm eff}>5000$~K have been mostly identified through their $v_r$.
Figure \ref{figure:mem_radec} shows the distribution of these in the surveyed area. The largest concentrations of new members are located on the ONC flanking fields, north of the ONC, and in southernmost end of L1641-south. This is not surprising, because in these areas there are no spectroscopic or X-ray surveys, and the sole membership indicator from previous literature studies is the IR-excess \citep{megeath2012}. Thus, it is expected that diskless members were not identified by this study. Similarly, new members are found on the edges of our surveyed area in L1641, because optical studies \citep{hsu2012,hsu2013,fang2009,fang2013}, as well as X-ray observations \citet{pillitteri2013} had a narrower field of view along the filament.

We must highlight that the sample of new candidate members is not unbiased, as it is originated from a method based on the previously known members, which is a heterogenous sample. Moreover, the method itself is not consistently reliable, as the ability to assign membership strongly depends on temperature. Thus, our membership estimation should be considered as a list of new ``candidates'' rather than a list of new ``confirmed members''
}

\section{Summary}
\label{section:summary}
Within the SDSS-III IN-SYNC survey, we have targeted the young stellar population of the Orion A molecular cloud with the APOGEE spectrograph. We have obtained high resolution $H-$band spectra for nearly 2700 individual sources, chosen among known members but also targeting sources of unknown membership. We extract stellar parameters ($T_{\rm eff}$, $\log g$, $v \sin i$, $A_V$) and radial velocities $v_r$ adopting a fitting procedure on grids of synthetic spectra. Consistent with the previous IN-SYNC surveys in the Perseus Cloud \citep{cottaar2014,cottaar2015,foster2015} we find that the stellar parameters extracted from the APOGEE spectra are in good agreement with literature values from optical studies, especially at low $T_{\rm eff}$ ($\lesssim 4000$K). Tables \ref{table:fitted_parameters} and \ref{table:additional_parameters} report our estimated stellar parameters, and membership estimates. Our main results can be summarized as follows:

\begin{itemize}
\item By comparing our $A_V$ estimates, derived for individual stars from the color excess $E(J-H)$ with respect to the intrinsic $(J-H)_0$ at a given $T_{\rm eff}$ with literature $A_V$ from optical spectroscopy, we find that a reddening law compatible with $R_V=5.5$ is better suited to reproduce the data than a diffuse ISM reddening law. Given that, at low $A_V$, $R_V=3.1$ was found to be adequate \citep{dario2010b}, we conclude that grain growth is present at increasing level of embeddedness, where in Orion also the volume density of gas increases.
\item We study the age spread throughout the region and find evidence for a large spread in stellar ages. We find a clear correlation between ages assigned from isochrones in the HRD and ages derived in the $\log g$--$T_{\rm eff}$ plane, indicating that, at least, a large spread in stellar radii is real. We also detect a clear trend between estimated ages and $A_V$, in that the oldest tail of the population is systematically unembedded. This may indicate progressive star formation along the line of sight, consistent with the fact that on-going star formation in the region, associated with the dense gas, is located behind the PMS population.
\item We study the foreground population candidate members, located in the region of NGC1980 and $\iota$ Ori proposed by \citet{alves2012} and \citet{bouy2014}. Based on our data, we confirm that this population is older than the rest of the population by about 40\% and less embedded. Moreover, in this declination range there is a significant underabundance of young ($<1Myr)$ stars, suggesting that star formation occurred here earlier than in the rest of the cloud. { However, we suggest that this population is not completely separate from the filament and that it would be unlikely for it to be located many pc in foreground,} given that the kinematic properties ($v_r$ and $\sigma_r$) appear identical to those of the younger, embedded members in the same region, and show no discontinuity with the kinematic gradient of the entire Orion A cloud.
\item Based on a combination of constraints considering stellar parameters ($T_{\rm eff}$, $\log L$, $\log g$), NIR photometry, and radial velocities, we identify 383 new candidate members. These are mostly located in areas of our FOV with no previous spectroscopic or X-ray coverage, and are mostly diskless PMS sources that could not be identified as members based on sole IR excess classifications. {\rm We stress that while statistically most of these new candidates should be members of the Orion A populations, our data do not allow us to confirm the membership for each individual source, and some candidate members may not be actual members.}
     The remaining sources likely to be field contaminants are predominantly intermediate temperature ($T_{\rm eff}>4500$K) background dwarfs; we also detect a clear background RGB sequence well evident in the $\log g$--$T_{\rm eff}$ plane.
\end{itemize}

\end{document}